\documentclass[prd,aps,showpacs,nofootinbib]{revtex4}
\usepackage{graphicx}
\usepackage{natbib}
\usepackage{bm}
\usepackage{amssymb}
\usepackage{amsmath}
\usepackage{euscript}
\usepackage{esint}
\usepackage{prelim2e}
\usepackage{slashed}

\begin{document}

\title{Large-time evolution of electron in photon bath}

\author{Kirill A. Kazakov}
\author{Vladimir V. Nikitin}
\affiliation{%
Department of Theoretical Physics, Physics Faculty, Moscow State
University, 119991, Moscow, Russian Federation}

\begin{abstract}
The problem of infrared divergence of the effective electromagnetic field produced by elementary charges is revisited using the model of an electron freely evolving in a photon bath. It is shown that for any finite travel time, the effective field of the electron is infrared-finite, and that in each order of perturbation theory the radiative contributions grow without bound in the large-time limit. Using the Schwinger-Keldysh formalism, factorization of divergent contributions in multi-loop diagrams is proved, and summation of the resulting infinite series is performed. It is demonstrated that the effective electromagnetic field of the electron vanishes in the large-time limit, and that this vanishing respects the total charge conservation and the Gauss law. It is concluded that the physical meaning of infrared singularity in the effective field is the existence of a peculiar irreversible spreading of electric charges, caused by their interaction with the photon field. This spreading exists in vacuum as well as at finite temperature, and shows itself in a damping of the off-diagonal elements of the momentum-space density matrix of electron, but does not affect its momentum probability distribution. It precludes preparation of spatially localized particle states at finite times by operating with free particle states in the remote past. Relationship of the obtained results to the Bloch-Nordsieck theorem is established, and discussed from the standpoint of measurability of the electromagnetic field. The effect of irreversible spreading on the electron diffraction in the classic two-slit experiment is determined, and is shown to be detectable in principle by modern devices already at room temperature.
\end{abstract}

\maketitle

\section{Introduction}

The low-energy behavior of quantum systems is an issue which is important in both fundamental and applied aspects of quantum field theory. Investigation of the low-energy properties of particle interactions constitutes an essential part in establishing the correspondence between classical and quantum theories, and finds numerous applications in all areas of quantum physics, from noise theory to quantum cosmology. As is well known, these properties essentially depend on the presence of massless particle states. In non-relativistic quantum mechanics, the existence of massless photons and gravitons shows itself in the form of long-range Coulomb and Newton forces between massive particles. Account of the radiation effects brings in, among other outcomes of the union of quantum theory with the relativistic principle, a factor that has no counterpart in the physics of systems involving only massive particles. Namely, the masslessness of photons and gravitons, together with the non-conservation of the particle number, imply that production of these quanta in any scattering process is beyond the experimental control. More precisely, finite sensitivity of any experimental setup does not allow one to distinguish scattering processes which involve different numbers of massless quanta with sufficiently small energy.

The role of this indistinguishability in the scattering theory is also known well. In the standard formulation using the S-matrix, the scattering process is considered formally on an infinite time interval, implying that the 4-momenta of virtual photons describing radiative corrections may take on arbitrarily small values. Integration over such momenta gives formally infinite results at every order of the perturbation theory. On the other hand, the uncontrollable production of photons with arbitrarily small energy means that the observed scattering cross-section is actually an infinite sum of terms, each of which represents the given scattering process with a definite number of extra real soft photons,  {\it i.e.,} photons with energies below the sensitivity threshold.\footnote{And with a  total energy going into unobserved photons less than the uncertainty in the energy of ``hard'' particles, {\it i.e.,} the particles being scattered.} Integration of the cross-section over the real soft photon momenta brings in another divergence which exactly cancels the divergence due to virtual photons, resolving thereby the infrared catastrophe of quantum electrodynamics \cite{bloch}. For brevity, this result will be referred to below as the Bloch-Nordsieck theorem. Similar cancelations take place in quantum gravity \cite{wein}, and in a more intricate way, in Yang-Mills theories \cite{leenaum}.

These results, though resolve the infrared catastrophe in a physically adequate way, leave open the question about possible physical manifestations of the infrared singularities, other than mere explicit dependence of the cross-sections on the sensitivity threshold. That the Bloch-Nordsieck theorem does not exhaust the infrared problem can be seen from the standpoint of the measurement theory. A result of fundamental importance regarding measurability of the electromagnetic field, proved by Bohr and Rosenfeld eighty years ago, asserts consistency of the principal limitations imposed by quantum theory on realizability of field measurements with formal predictions of quantum electrodynamics \cite{bohr1}. More specifically, all statistical predictions following from the formal relations between electromagnetic field operators can be verified, with accuracy limited only by the uncertainty principle, using an appropriately designed {\it macroscopic} measuring device. The demonstration given by Bohr and Rosenfeld refuted objections against the possibility of such verification, raised earlier by Landau and Peierls \cite{landau}, which were based on consideration of the field measurement using {\it single} test charge. It is existence of uncontrollable radiation by the test charge under the influence of the field being measured that led the authors of \cite{landau} to the conclusion that quantum electrodynamics imposes additional limitations on the accuracy of field measurements, which turned out to be significantly more stringent than those following from the uncertainty principle. In particular, the use of single test charge restricts the accuracy of separate measurements of individual field components, in contradiction with the formal apparatus of quantum electrodynamics, which implies no such restriction. An important conclusion of the work \cite{bohr1} is that despite the principal impossibility to control the radiation produced by test bodies, the effect of this radiation on their motion can be compensated with the help of an appropriate experimental arrangement, but such compensation is possible only when the test bodies employed consist of sufficiently many elementary charges.\footnote{See also \cite{bohr2}. This result was carried over to the case of gravitational interaction by DeWitt \cite{dewitt}.}

Suppose now that we want to determine the electromagnetic field produced by a free electron in a given state. It follows from what was just said that scattering a test charge on the electron is not the best choice for this purpose. From the theoretical point of view, this means that using the S-matrix (say, the two-particle scattering amplitude) to describe the electron field generally is not adequate, as it unavoidably misses part of information about this field, namely that hidden by the uncontrollable radiation. Instead, since all statistical properties of the electromagnetic field are encoded in its operator, a complete description of the electron field can be obtained in terms of expectation values of the products of field operators, evaluated over the given state. Then the results of \cite{bohr1} guarantee that the predictions obtained in this way are amenable to experimental verification. One of the most important quantities of this sort is the expectation value of the electromagnetic field itself, called also mean, or effective, field. Thus, the very definition of the effective field excludes from consideration the uncontrollable radiation produced by the test bodies (whereas the uncontrollable radiation from the system that produces the field being measured is fully taken into account by the effective field formalism, Cf. Secs.~\ref{prelim}, \ref{rules}).

The purpose of the present paper is to investigate infrared properties of the effective electromagnetic field in the case when the field-producing electron is embedded in a photon bath at finite temperature. At zero temperature, this problem was considered already at the dawn of quantum field theory, to determine radiative corrections to the field of a classical point source ({\it e.g.,} atomic nucleus) \cite{serber}. The restriction to classical, that is, sufficiently heavy source is necessary in the conventional formulation precisely because of the presence of infrared divergences: Radiative corrections to the electromagnetic formfactors of charged particles vanish in the large-mass limit, which gives a formal reason to put the question about their divergence aside. Another reason which is often used in the literature to discard infrared-divergent contributions to the effective field is that at every order of the perturbation theory, such contributions are local, in the sense that they vanish when considered within the long-range expansion with respect to the distance from the source.\footnote{This is actually the only reason to get rid of infrared-divergent contributions to the effective gravitational field, for the radiative corrections to the gravitational formfactors do not vanish in the large-mass limit.} It is worth mentioning in this connection that despite numerous attempts \cite{tryon,weldon,indumathi,manjavidze,muller}, extension of the S-matrix formalism to finite temperatures is still an open question. Therefore, regardless of the fundamental reasons given above, the effective field is an indispensable means for studying temperature effects in quantum field theory.

The following circumstance is crucial for the discussion of infrared singularity in the effective field. The infrared divergences occur because evolution of the field-producing system is considered on an infinite time interval. An infinite temporal extent is required already by the procedure of adiabatic switching of the interaction, which is widely employed in quantum field calculations, in particular, in constructing the scattering matrix. The conclusion to be drawn from this fact is that in the presence of massless particles, special justification is required to use the notion of infinitely remote past. The Bloch-Nordsieck theorem gives such justification in the case of the S-matrix, but in the effective field formalism this notion turns out to be physically inadequate. The point is that, as was demonstrated in Ref.~\cite{jpa}, the infrared singularity in the effective field signifies existence of a peculiar spreading of the source particle, which precludes preparation of a spatially localized particle state at finite times by operating with arbitrary free particle states in the remote past. This was shown by evaluating the effective electromagnetic field of an electron, regularized by means of the momentum cutoff method appropriately modified to allow factorization of the infrared contributions in multi-loop diagrams (called $\lambda$-regularization in Ref.~\cite{jpa}). The effective field was found to vanish at any given spatial point in the limit of removed regularization, in a way that respects the total charge conservation. It was also argued that the momentum cutoff can be endowed with a physical meaning as estimating the inverse duration of the measurement process, which made it possible to explicitly describe the electron evolution subjected to the irreversible spreading, and to estimate its possible observational effects.

Except for a modification of the regularization scheme, investigation carried out in Ref.~\cite{jpa} employs the conventional method of calculating the effective field, based on consideration of the system on an infinite time interval. This investigation is therefore incomplete in view of what has been said regarding the role of the temporal extent in studying the infrared singularity. In addition to that, the use of an auxiliary infrared regularization raises the question about scheme dependence of the obtained results. At last, it is necessary to establish an exact form of the time-dependence of the electron state affected by the infrared singularity. Thus, we have to consider the electron evolution on a finite time interval, and to determine the leading large-time contributions to the electron density matrix. Finiteness of the time interval renders Feynman integrals infrared-convergent, removing thereby the question of scheme dependence, as there is no need in auxiliary infrared regularization. The infrared singularity is now contained in the large-time asymptotic of the effective field, or equivalently, of the effective electromagnetic current of the electron, and the problem is to consistently extract this asymptotic. It turns out that this requires significant modification of the standard calculational scheme, because restricting the consideration to a finite interval raises the issue of initial conditions for the electromagnetic field, which in turn enforces using the canonical Coulomb gauge instead of the covariant one, to avoid violation of the Gauss law in the initial state. It will be shown below how these issues are interrelated, and why they do not arise in the standard formulation based on the adiabatic switching of the interaction.

The paper is organized as follows. The main tools to be used to study electron evolution in a photon bath are described in Sec.~\ref{prelim}. Here we also identify two time scales characterizing two essentially different stages of the electron evolution -- the infrared thermalization corresponding to the infrared singularity, and the usual kinetic relaxation of the electron momentum, described by a kinetic equation. The complications introduced by finiteness of the time interval are discussed in detail in Sec.~\ref{form}. It is shown, in particular, that the standard procedure of transition from a canonical gauge to the Lorentz gauge cannot be accomplished in the usual way: the gauge-non-invariance of the electron density matrix leads to appearance of a Lorentz-non-invariant term in the Lagrangian. The resulting modification of the Feynman rules in the Schwinger-Keldysh method is described in Sec.~\ref{perturbth}. The infrared thermalization is studied in Sec.~\ref{infrared_calc}. First of all, the set of diagrams contributing to the large-time asymptotic of the effective current is identified in Sec.~\ref{approx}; this set turns out to be different from that representing the effective current within the $\lambda$-regularization when the initial electron state is specified at $t=-\infty.$ Factorization and summation of the infrared contributions are then performed in Sec.~\ref{factorization}. Section~\ref{effmat_calc} contains an alternative derivation of the main result, which uses specifics of the four-dimensional Feynman integrals to reduce the problem of extracting the large-time asymptotic to solving a differential equation for the electron density matrix. Some applications are given in Sec.~\ref{phexamples}: The physical meaning of the infrared singularity as representing an irreversible spreading of electric charges is illustrated in Sec.~\ref{gausswave} by working out evolution of a Gaussian wave-packet. The effect of infrared thermalization on the electron diffraction in the classic two-slit experiment is determined in Sec.~\ref{diffraction}, interpreted in terms of decoherence, and is shown to be detectable in a proper experimental arrangement already at room temperature. Finally, it is demonstrated in Sec.~\ref{relaxation} how interaction of the electron with non-infrared photons leads to the usual relaxation of the electron momentum. This is shown by using the method of Sec.~\ref{effmat_calc} to obtain a differential equation for the electron momentum distribution, which turns out to be the usual kinetic equation. Conclusions are drawn in Sec.~\ref{conclusions}. The paper has an appendix containing discussion of the role of the non-invariant term in the Lagrangian in Lorentz gauge.

\section{General formulation}\label{generalformulation}

\subsection{The model}\label{prelim}

Consider a non-relativistic electron of mass $m,$ interacting with virtual and real photons in equilibrium at finite temperature\footnote{We use relativistic units $\hbar = c = 1.$ Also, the Minkowski metric is $\eta_{\mu\nu} = {\rm diag}\{+1,-1,-1,-1\}.$} $T \ll m.$ We shall deal with two closely related physical quantities: the effective (mean) electromagnetic current of the electron, $J^{\rm eff}_\mu,$ and the effective electromagnetic field produced, $F^{\rm eff}_{\mu\nu}.$
They are defined at every space-time point $x$ as the expectation values
$$F^{\rm eff}_{\mu\nu}(x) = \left\langle F_{\mu\nu}[A(x)]\right\rangle, \quad
J^{\rm eff}_\mu(x)=\left\langle J_\mu(x) \right\rangle,
$$
where $A_\mu (x),$ $J^\mu(x)={\bar \psi}(x)\gamma^{\mu}\psi(x)$ are Heisenberg picture operators of the electromagnetic potential and current, respectively, and angular brackets denote averaging over given (mixed) state of the electron and photons. This definition applies to systems at finite temperatures as well as at $T=0,$ and according to the general rules of quantum theory, it yields physical quantities averaged over series of measurements in a given state. Since the electromagnetic field strength is linear in $A_{\mu},$ one has $F^{\rm eff}_{\mu\nu}(x)=\partial_{\mu}A^{\rm eff}_{\nu}(x)-\partial_{\nu}A^{\rm eff}_{\mu}(x),$ where $A^{\rm eff}_{\mu}(x) = \left\langle A_{\mu}(x)\right\rangle$ is the effective electromagnetic potential. Also, the linearity of field equations implies that
\begin{eqnarray}\label{HeisAver}
\partial_\mu F^{\rm eff}_{\mu\nu}(x) = eJ^{\rm eff}_\nu(x)
\end{eqnarray}\noindent
Before we proceed to detailed calculation of the effective quantities introduced, it is useful to express them via the electron density matrix, which helps exposing a certain complementarity between the effective field and another important object -- transition probability. Any state vector of the electron interacting with photons can be expanded in the direct products of states describing a free electron, $|e\rangle,$ and states describing arbitrary number of free photons, $|\phi\rangle,$ all with definite momenta and spin/polarizations. Consider an electron which is in a pure state $|\psi_0\rangle$ at the initial time instant $t_0,$ whereas photons are in equilibrium. This system is described by the density matrix $\sum_{\phi_0}w(\phi_0)|\psi_0\rangle|\phi_0\rangle \langle \phi_0|\langle \psi_0|,$ where $w(\phi) = {\rm e}^{-E_{\phi}/T}/\sum_{\phi}{\rm e}^{-E_{\phi}/T},$ $E_{\phi}$ being the energy of photons in a state $|\phi \rangle,$ and summation is over all such states. The probability to find the system in a state $|e\rangle|\phi\rangle$ at time $t> t_0$ is
\begin{eqnarray}\nonumber
\sum_{\phi_0} w(\phi_0)\langle e|\langle\phi|U(t,t_0)|\psi_0\rangle|\phi_0\rangle\langle \phi_0|\langle \psi_0|U(t_0,t)|e\rangle|\phi\rangle\,,
\end{eqnarray}\noindent where $U(t,t_0)$ is the evolution operator on the interval $(t_0,t).$ Being interested solely in the electron evolution, we sum over all final photon states. Thus, the probability of transition of the electron into a state $|e\rangle$ takes the form
\begin{eqnarray}\label{diag}
\sum_{\phi,\phi_0} w(\phi_0)\langle e|\langle\phi|U(t,t_0)|\psi_0\rangle|\phi_0\rangle\langle \phi_0|\langle \psi_0|U(t_0,t)|e\rangle|\phi\rangle = \langle e|\varrho(t)|e\rangle,
\end{eqnarray}\noindent
where
\begin{eqnarray}\label{densitymatrix}
\varrho(t)=\sum_{\phi,\phi_0}w(\phi_0) \langle  \phi|U(t,t_0)|\psi_0\rangle|\phi_0\rangle\langle \phi_0|\langle \psi_0|U(t_0,t)|\phi\rangle.
\end{eqnarray}\noindent Since creation of electron-positron pairs is negligible under condition $T\ll m,$ the initial state vector $|\psi_0\rangle|\phi_0\rangle$ is carried by $U(t,t_0)$ into a one-electron state. Expression (\ref{densitymatrix}) is thus  nothing but the electron density matrix at time $t.$ It is to be noted that as defined, this matrix takes into account all radiation effects, including those due to soft photons emitted by the electron (uncontrollable radiation). Indeed, summation in Eq.~(\ref{densitymatrix}) is over all possible photon states $|\phi\rangle.$ Now consider the mean value problem, and let $F$ be an operator built of the fermion fields only ({\it e.g.,} the electromagnetic current). Then one has $F|e\rangle|\phi\rangle=|\phi\rangle F|e\rangle,$ and using completeness of the products $|e\rangle|\phi\rangle$ the effective value of $F$ at time $t$ can be written as
\begin{eqnarray}\nonumber
F^{\rm eff}(t) &=& \sum_{\phi_0}w(\phi_0)\langle \phi_0|\langle \psi_0|U(t_0,t)F U(t,t_0)|\psi_0\rangle|\phi_0\rangle \nonumber\\
&=& \sum_{\phi,\phi_0,e}w(\phi_0)\langle \phi_0|\langle \psi_0|U(t_0,t)F |e\rangle|\phi\rangle\langle \phi|\langle e|U(t,t_0)|\psi_0\rangle|\phi_0\rangle \nonumber\\
&=& \sum_e\langle e|\varrho(t)F|e\rangle.\label{feff}
\end{eqnarray}\noindent
It is seen that the inclusive transition probability and the effective quantities are expressed via the same object -- the electron density matrix. An important difference is that in the first instance we deal only with diagonal elements of the density matrix, whereas the effective fields depend also on off-diagonal elements. From the point of view of the infrared problem, this means that the cancelation of infrared singularities in the S-matrix, asserted by the Bloch-Nordsieck theorem and its generalizations to $T\ne 0$ \cite{tryon,weldon,indumathi}, implies finiteness of only diagonal elements of the electron density matrix in the limit $t_0\to-\infty,$ $t\to + \infty.$ In other words, consideration of the effective field gives an important piece of information about the system, which is not contained in the S-matrix. In fact, the existence of an irreversible charge spreading, revealed by consideration of the effective electromagnetic field, is related precisely to the off-diagonal elements of the electron density matrix.

In the subsequent study of the effective electromagnetic field of electron, a formally more general setting will be useful wherein at $t_0$ the electron is already in a mixed state described by the density matrix $\varrho_0,$ and $F$ is an arbitrary gauge invariant operator built of fermion as well as electromagnetic field operators. Equation~(\ref{feff}) is then replaced by
\begin{eqnarray}\label{eff1}
F^{\rm eff}(t) &=& {\rm Tr}_e\sum_{\phi_0}w(\phi_0)\langle \phi_0|\varrho_0U(t_0,t)F U(t,t_0)|\phi_0\rangle \nonumber\\
&=& \EuScript{N}^{-1}{\rm Tr}_e{\rm Tr}_{\phi}\left(U(t_0,t)F U(t,t_0)\varrho_0 {\rm e}^{- \beta H_\phi }\right)\nonumber\\
&&\EuScript{N} \equiv {\rm Tr}_{\phi}\left( {\rm e}^{- \beta H_\phi }\right), \quad \beta \equiv 1/T \,.
\end{eqnarray}
\noindent where $H_\phi$ is the Hamiltonian of free photons, and ${\rm Tr}_{\phi},$ ${\rm Tr}_e,$ denote traces over all photon states and the single electron states, respectively. This expression is merely a longwinded definition $F^{\rm eff}(t)=\left\langle F(t) \right\rangle$ specified to the given initial condition.

Our goal is to infer physical consequences of the infrared singularity in the effective electromagnetic field of the electron, or equivalently, in its effective current. Accordingly, we will be interested in the large-time asymptotics of these quantities. There are two essentially different types of contributions which diverge for $t \to \infty,$ when treated within the perturbation theory. They describe different stages of the relaxation process in the system, which are characterized by significantly different time scales. Namely, the faster process is the infrared thermalization described in \cite{jpa} within the $\lambda$-regularization, which takes place in vacuum as well as at finite temperature, and which shows itself in a damping of the off-diagonal elements of the momentum-space density matrix of the electron. Restoring for a moment the ordinary units, Ref.~\cite{jpa} gives the following estimate for the characteristic time of this process
\begin{eqnarray}\nonumber
\tau_1 \sim \frac{m^2 c^2 r^2}{\hbar\alpha T}\,,
\end{eqnarray}
where $\alpha = e^2/(4\pi\hbar c)$ is the fine-structure constant, and $r$ is the characteristic length of the problem, {\it e.g.,} the distance between the electron and the point of observation of its field, or the spacing of an interference pattern in the two-slit experiment (see Sec.~\ref{diffraction}), {\it etc.} Consideration of this process is the main subject of the present paper (subsequent calculations confirm the above estimate for $\tau_1$).

The slower process is the usual relaxation of the electron momentum, caused by its collisions with thermal photons, which occurs only at $T\ne 0.$ Its rate can be estimated by noting that the cross-section of the low-energy electron-photon scattering is of order $(\alpha\hbar/m c)^2,$ whereas the photon density is $\sim (T/\hbar c)^3.$ Hence, the electron mean free time is
\begin{eqnarray}\nonumber
\tau_2 \sim \frac{\hbar m^2 c^4}{\alpha^2 T^3}\,.
\end{eqnarray}\noindent
This process is described by the usual kinetic equation for the electron momentum probability distribution (that is, the diagonal elements of the electron density matrix), and will be considered in Sec.~\ref{relaxation}.

The ratio of the two time scales,
\begin{eqnarray}\label{tauratio}
\frac{\tau_1}{\tau_2} \sim \alpha \left(\frac{rT}{\hbar c}\right)^2,
\end{eqnarray}\noindent is to be considered small within the perturbation theory. Less formally, $$\frac{\tau_1}{\tau_2} \sim \frac{1}{137}(4.36rT)^2,$$ where $T$ is assumed to be expressed in kelvins, and $r$ in centimeters, so that the ratio is small in microscopic processes ($r\sim 10^{-8}$\,cm) for all practically important temperatures.

Below, the two stages of the electron evolution are considered separately, the reason being that, as was already mentioned above, they deal with different elements of the electron density matrix.

\subsection{Interaction picture and gauge conditions}\label{form}

\begin{figure}[h]
\begin{center}
     \includegraphics{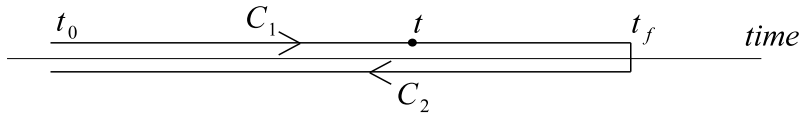}
\caption{Time-integration contour in Eq.~(\ref{eff2}).}
\label{contour}
\end{center}
\end{figure}

To evaluate $F^{\rm eff}$, we shall use the Schwinger-Keldysh formalism \cite{schwinger,keldysh}, according to which Eq.~(\ref{eff1}) can be written in the interaction picture as
\begin{eqnarray}\label{eff2}
F^{\rm eff}(t) = \EuScript{N}^{-1}{\rm Tr_\phi}{\rm Tr_e} \left(  T_c \left[ {\exp {\rm i}\left( {\int\limits_C {{\rm d}^4 y} L_I (y)} \right) F (t)} \right]{\rm e}^{- \beta H_\phi } \varrho_0\right)\,,
\end{eqnarray}
where $L_I$ is the interaction Lagrangian, $y^0$-integration is along the time-contour $C=C_1\cup C_2$ running from $t_0$ to $t_f\geq t$ and back as shown in Fig.~\ref{contour}, and $T_c$ denotes operator ordering along this contour. It is conventional to take the limit $t_f \to \infty$ in this formula, but the choice $t_f=t$ is more appropriate for our purposes.

To completely specify the scheme of calculations, one has to fix gauge invariance of the theory, which can be done by including a gauge-fixing term into Lagrangian density in Eq.~(\ref{eff2}). The amount of computational labor essentially depends on the gauge choice, and covariant gauges are well-known to be of great advantage over canonical ones in this respect, but their use requires special justification. A mere reference to gauge-independence of the effective electromagnetic current is not sufficient: One has to prove that the use of covariant rules gives the same results as the original canonical method. It turns out that Eq.~(\ref{eff2}) is the case where covariant techniques is not applicable, the failure being directly related to finiteness of the initial instant $t_0.$ To explain the point, it will be convenient to follow the general Faddeev-Popov method \cite{slavnov} of transition to covariant gauge, despite the fact that the gauge group in the present case is only Abelian.

We choose to start with the Coulomb gauge (details of quantization in this gauge can be found in \cite{tyutin,Weinberg})
$$\partial_i A^i=0, \quad i=1,2,3.$$ This gauge admits canonical quantization, whereby the gauge condition can be considered as an operator relation. Expression (\ref{eff2}) written in terms of the functional integral takes the form
\begin{eqnarray}\label{eff_cont}
F^{\rm eff}(t) = \EuScript{N}^{-1}
\int{\rm d}\psi{\rm d}\bar{\psi}
\langle \psi^1|\varrho_0|\psi^2\rangle \int {\rm d} A_\nu
\langle A^1_i|{\rm e}^{- \beta H_\phi}|A^2_i\rangle
 B_C\delta(\partial_i A_i)
\nonumber \\\times
\exp {\rm i}\left( \int\limits_{C}{{\rm d}^4 y} L(y) \right)F(t).
\end{eqnarray}\noindent
Here ${\rm d}\psi{\rm d}\bar{\psi}$ is shorthand for the fermion functional integral measure, $$\prod\limits_{\bm{x},t'\in C_1}{\rm d} \psi^1(\bm{x},t'){\rm d} \bar\psi^1(\bm{x},t') \prod\limits_{\bm{x},t'\in C_2}{\rm d}\psi^2(\bm{x},t') {\rm d}\bar\psi^2(\bm{x},t'),$$ and similarly for the $A$-field; the superscripts $1,2$ distinguish fields belonging to the corresponding branch of the time contour $C;$ integration is over all fields satisfying $\psi^1(t)=\psi^2(t),$ $\bar\psi^1(t)=\bar\psi^2(t),$ $A^1(t)=A^2(t).$ Next, $|\psi^{1,2}\rangle,$ $|A^{1,2}_i\rangle$ are eigenvectors of the field operators at $t_0,$ {\it e.g.,} ${\hat{A}_i(t_0,\bm{x})|A^1_i\rangle=A^1_i(t_0,\bm{x})|A^1_i\rangle},$ {\it etc.} Finally, the factor $ B_C$ (a formally infinite constant) is defined by
\begin{eqnarray}\label{D_C}
 B_C\int {\rm d}\omega \delta(\partial_i A^{\omega}_i)=1,
\end{eqnarray}\noindent
where integration is over the gauge group, and $A^{\omega}_\nu$ is the result of action of the gauge group element $\omega$ on the field $A_\nu.$
Note that $|A_i\rangle$ is independent of $A_0,$ since $A_0$ is not a dynamical variable.  Hence, integration over $A_0$ yields the Coulomb law
\begin{eqnarray}\label{Coul_Gauss}
A_0=-\triangle^{-1} J_0,
\end{eqnarray}
which is thus also treated as an operator relation. Now, let us define $ B_L$ (another infinite constant) by
$$ B_L\int {\rm d}\omega \delta(\partial^\mu A^{\omega}_\mu(x)-a(x))=1,$$
with $a(x)$ an arbitrary function of spacetime coordinates, and multiply (\ref{eff_cont}) by unity:
\begin{eqnarray}\label{eff_cont1}
F^{\rm eff}(t) = \EuScript{N}^{-1}
\int{\rm d}\psi{\rm d}\bar{\psi}&&\hspace{-0,3cm}
\langle \psi^1|\varrho_0|\psi^2\rangle
\int{\rm d} A_\nu
\langle A^1_i|{\rm e}^{- \beta H_\phi}|A^2_i\rangle
 B_C\delta(\partial_i A_i)
\nonumber \\&&\times
 B_L\int {\rm d}\omega\delta(\partial^\mu A^{\omega}_\mu-a)
\exp {\rm i}\left(\int\limits_C {{\rm d}^4 y} L(y)\right)F(t).
\end{eqnarray}\noindent
A change of integration variables $\psi^\omega \to \psi,$ $\bar\psi^\omega \to \bar{\psi},$ $A^\omega \to A,$ $\omega \to \omega^{-1}$ gives
\begin{eqnarray}
F^{\rm eff}(t) = \EuScript{N}^{-1}
\int {\rm d}\omega
\int {\rm d}\psi{\rm d}\bar{\psi}&&\hspace{-0,3cm}
\langle \psi^{1\omega}|\varrho_0|\psi^{2\omega}\rangle
\int{\rm d} A_\nu\langle A^{1\omega}_i|{\rm e}^{- \beta H_\phi}|A^{2\omega}_i\rangle
 B_C\delta(\partial_i A^{\omega}_i)
\nonumber \\&&\times
 B_L \delta(\partial^\mu A_\mu-a)
\exp {\rm i}\left(\int\limits_C {{\rm d}^4 y} L(y)\right)F(t,\bm{x}),\nonumber
\end{eqnarray}\noindent
where gauge invariance of the Lagrangian $L$ and of the operator $F$ was taken into account. Moreover, since the photon state vectors are defined by transversal components of $\bm{A},$ they are gauge-invariant, $|A^{1,2\omega}_i\rangle=|A^{1,2}_i\rangle.$ But this is not true of the fermion states $|\psi^{1,2}\rangle.$ Therefore, $\langle \psi^{1\omega}|\varrho_0|\psi^{2\omega}\rangle \ne \langle \psi^{1}|\varrho_0|\psi^{2}\rangle,$ for $\varrho_0$ is not gauge-invariant.
Therefore, integration over $\omega$ leads to the following expression for the effective current in the Lorentz gauge
\begin{eqnarray}
F^{\rm eff}(t) = \EuScript{N}^{-1}
\int {\rm d}\psi{\rm d}\bar{\psi}\int{\rm d} A_\nu &&\hspace{-0,3cm}
\langle {\rm e}^{{\rm i}e \alpha(A)}\psi^1|\varrho_0|{\rm e}^{{\rm i}e \alpha(A)}
\psi^2\rangle
\langle A^1_i|{\rm e}^{- \beta H_\phi}|A^2_i\rangle
\nonumber \\&&\times
 B_L \delta(\partial^\mu A_\mu - a)
\exp {\rm i}\left(\int\limits_C {{\rm d}^4 y} L(y)\right)F(t),\nonumber
\end{eqnarray}\noindent where $\alpha(A) = -\triangle^{-1}\partial_iA_i$. If the limit $t_0 \to -\infty$ is taken, and the interaction is negligible in the remote past, the function $\langle {\rm e}^{{\rm i}e \alpha(A)}\psi^1|\varrho_0|{\rm e}^{{\rm i}e \alpha(A)} \psi^2\rangle$ can be replaced by $\langle \psi^1|\varrho_0|\psi^2\rangle,$ for then the interaction can be adiabatically switched off in the remote past,
$$e \to 0, \quad t \to - \infty,$$ which kills the phase factor ${\rm e}^{{\rm i}e \alpha(A)}.$ This is exactly what happens in the $S$-matrix formalism, and is realized there through the use of in- and out-states. At finite $t_0,$ one can get rid of the phase factor by making the inverse change of the fermion integration variables $\psi\to{\rm e}^{-{\rm i}e \alpha(A)}\psi,$ $\bar\psi\to{\rm e}^{{\rm i}e \alpha(A)}\bar\psi,$ but then an additional term appears in the Lagrangian
\begin{eqnarray}
F^{\rm eff}(t) = &&\hspace{-0,3cm}\EuScript{N}^{-1}
\int {\rm d}\psi{\rm d}\bar{\psi}\int{\rm d} A_\nu
\langle \psi^1|\varrho_0|\psi^2\rangle
\langle A^1_i|{\rm e}^{- \beta H_\phi}|A^2_i\rangle
\nonumber \\&&\times
 B_L \delta(\partial^\mu A_\mu-a)
\exp {\rm i}\left(\int\limits_C {{\rm d}^4 y}
\left[L(y)-e\bar\psi\gamma^\mu\psi\triangle^{-1}\partial_\mu\partial_iA_i\right]
\right)F(t).\nonumber
\end{eqnarray}\noindent
Finally, independence of $a$ allows one to average this equation with a weight $\exp{\left(-{{\rm i}\int d^4 y a^2(y)}/2\xi\right)},$ $\xi={\rm const},$ to obtain
\begin{eqnarray}\label{eff_cont3}
F^{\rm eff}(t) = \EuScript{N}^{-1}
\int {\rm d}\psi{\rm d}\bar{\psi}\int{\rm d} A_\nu &&\hspace{-0,3cm}
\langle \psi^1|\varrho_0|\psi^2\rangle
\langle A^1_i|{\rm e}^{- \beta H_\phi}|A^2_i\rangle
\nonumber \\&&\times
 B_L
\exp {\rm i}\left(\int\limits_C {{\rm d}^4 y} L_l(y)\right)F(t),
\end{eqnarray}\noindent
where
\begin{eqnarray}\label{L_l}
L_l=L-{\left(\partial^\mu A_\mu\right)^2}/{2\xi}-e J^\mu\triangle^{-1}\partial_\mu\partial_iA_i.
\end{eqnarray}\noindent The averaging affects only the normalization factor denoted $\EuScript{N},$ as before. The Lagrangian $L_l$ generates Feynman rules to be used to compute the effective field in the Lorentz gauge. It is seen that in contrast to the S-matrix formalism (based on taking the limit $t_0 \to - \infty,$ $t_f \to + \infty$), these rules are not Lorentz-covariant: Explicit Lorentz invariance is broken by the last term in Eq.~(\ref{L_l}), which largely deprives Lorentz gauge of its advantage over the Coulomb gauge. That the non-invariant term in $L_l$ cannot be discarded is demonstrated in Appendix A where it is shown that doing so would violate the Gauss law already in the tree approximation.\footnote{Note that the charge conservation $\partial_{\mu}J^{\mu}=0$ does not entail vanishing of the non-invariant contribution: Integration by parts in Eq.~(\ref{eff_cont3}) yields an integral of $e(J^1_0 - J^2_0)\triangle^{-1}\partial_iA_i$ over the hyperplane $y^0=t_0.$}

At last, it is worth mentioning that
\begin{eqnarray}\label{lorentzav}
\partial^\mu A^{\rm eff}_\mu(x) = {\rm const}.
\end{eqnarray}\noindent Indeed, it follows from Eq.~(\ref{eff_cont3}) that $A^{\rm eff}_\mu$ satisfies
\begin{eqnarray}
\Box A^{\rm eff}_\mu(x) - \left(1 - \frac{1}{\xi}\right)\partial_{\mu}\partial^{\nu}A^{\rm eff}_\nu(x) = J^{\rm eff}_\mu(x). \nonumber
\end{eqnarray}\noindent On the other hand, Eq.~(\ref{HeisAver}) is still in force, implying that $\partial_{\mu}\partial^{\nu}A^{\rm eff}_\nu(x)=0.$

\section{Perturbation theory}\label{perturbth}

\subsection{Feynman rules}\label{rules}

Perturbation expansion of Eq.~(\ref{eff2}) generates expressions of the form
\begin{eqnarray}\nonumber
{\rm Tr_\phi}{\rm Tr_e} \left(  T_c \left[ A_\mu(x_1)J^\mu(x_1)A_\nu(x_2)J^\nu(x_2)\cdots\right]{\rm e}^{- \beta H_\phi } \varrho_0\right)
\end{eqnarray}\noindent which on account of commutativity of the interaction picture operators factorize to
\begin{eqnarray}\nonumber
{\rm Tr_\phi}\left(  T_c \left[ A_\mu(x_1)A_\nu(x_2)\cdots\right]{\rm e}^{- \beta H_\phi } \right)
{\rm Tr_e}\left(  T_c \left[ J^\mu(x_1)J^\nu(x_2)\cdots\right]\varrho_0\right).
\end{eqnarray}\noindent The resulting Green functions can be further expanded in the products of particle propagators using the real-time techniques \cite{landsman1991,niemi}. As usual, the $T_c$-ordering promotes the field propagators into $2\times2$ matrices, {\it e.g.,} $$D^{(ij)}(x-y) = {\rm i}{\rm Tr_\phi}\left(  T_c \left[ A^{(i)}_\mu(x)A^{(j)}_\nu(y)\right]{\rm e}^{- \beta H_\phi }\right)$$ for the electromagnetic field, and similarly for the fermion field; the matrix indices $i,j$ take the value $1(2)$ for fields on the forward (backward) branch of the contour $C.$ In momentum space, the electron propagator has the form
\begin{eqnarray}\label{electronpr}
D^{(11)}(q) &=& - \tilde D^{(22)}(q)=\frac{\slashed{q} + m}{m^2-q^2-{\rm i}0}, \nonumber\\
D^{(21)}(q) &=& 2\pi{\rm i}\theta(q_0)\delta(m^2-q^2)(\slashed{q} + m), \nonumber\\
D^{(12)}(q) &=& 2\pi{\rm i}\theta(-q_0)\delta(m^2-q^2)(\slashed{q} + m),
\end{eqnarray}\noindent
where the tilde symbolizes an operation of complex conjugation with respect to which the Dirac matrices are real. The photon propagator reads
\begin{eqnarray}\label{photonpr}
D^{(11)}_{\mu\nu}(k) &=& - \tilde D^{(22)}(k) =
\left[\frac{1}{k^2+{\rm i}0}-2\pi{\rm i}n(\bm{k})\delta(k^2)\right]d_{\mu\nu}(k),
\nonumber\\
D^{(21)}_{\mu\nu}(k) &=& - 2\pi{\rm i}\left[\theta(k_0)+n(\bm{k})\right]\delta(k^2)d_{\mu\nu}(k), \nonumber\\
D^{(12)}_{\mu\nu}(k) &=& - 2\pi{\rm i}\left[\theta(-k_0)+n(\bm{k})\right]\delta(k^2)d_{\mu\nu}(k), \quad n(\bm{k})=\frac{1}{{\rm e}^{\beta|\bm{k}|} - 1}\,,\label{nk}
\end{eqnarray}\noindent
where
\begin{eqnarray}\label{coulombd}
d_{\mu\nu}(k)=\eta_{\mu\nu}-\frac{k_0 k_\mu \eta_\nu+k_0 k_\nu \eta_\mu-k_\mu k_\nu}{\bm{k}^2}\,, \quad \eta^{\mu} = (1,\bm{0})
\end{eqnarray}\noindent in the Coulomb gauge, whereas in the Lorentz gauge
\begin{eqnarray}
d_{\mu\nu}(k)=\eta_{\mu\nu}+(\xi-1)\frac{k_\mu k_\nu}{k^2}.
\end{eqnarray}\noindent
As usual, the interaction vertices are generated by the Lagrangian $L_I,$ and are assigned indices 1 or 2, depending on the branch of the contour $C$ to which the given vertex belongs, with an extra factor $(-1)$ for each type-2 vertex. The point of observation $x$ is assigned index 1. Each propagator connects vertices of the types assigned to the propagator ends, and integration in the vertices is over all space and over the time interval $(t_0,t).$ Diagrammatically, the electron and photon propagators will be depicted by straight and wavy lines, respectively. The operators being averaged ($A_{\mu}(x),$ $J^{\mu}(x),$ {\it etc.}) will be collectively denoted by an open circle, while the interaction vertices by full circles.

\begin{figure}[h]
\begin{center}
     \includegraphics{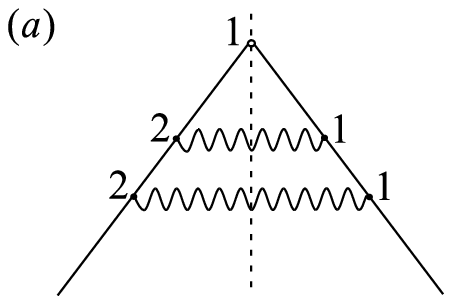}
     \includegraphics{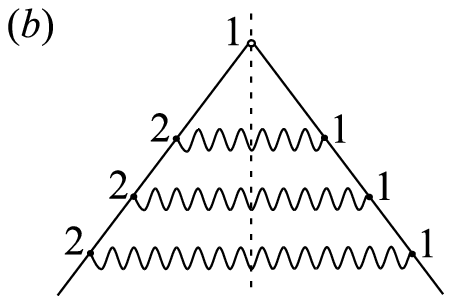}
\caption{Diagrams contributing to the effective electron current. (a) A fourth-order diagram describing the effect of electron-photon scattering. (b) A sixth-order diagram describing the effect of electron-photon scattering accompanied by a photon emisson by the electron. Vertical broken lines show unitarity cuts to be made to relate these diagrams to the scattering cross-sections.}
\label{example}
\end{center}
\end{figure}

The diagrammatic representation of the effective field in the Schwinger-Keldysh technique has its specifics when compared to the usual S-matrix diagrammatics. As long as the single electron problem is considered, this primarily concerns representation of the photons present in the system. The effect of the heat bath photons is taken into account by the term proportional to $n(\bm{k})$ in Eq.~(\ref{photonpr}). As to the photons produced by the electron, their effect is also encoded in the internal photon lines, in contrast to the S-matrix case where photons emitted by charges (in particular, the uncontrollable radiation) are represented by external photon lines attached to the charged particle propagators. It is integration of the scattering cross-sections over momenta of these photons that cancels the infrared divergences due to virtual photons appearing in the loops. The difference in the graphical representation arises because in the S-matrix case, the  photons emitted by charges are not present in the {\it in}-state, and appear only in the {\it out}-state as a result of the scattering, whereas the effective field is evaluated entirely over the given {\it in}-state, and no {\it out}-state ever appears in the formalism. In this respect, the effective field diagrammatics bears some resemblance with that of the scattering cross-sections transformed using the unitarity relations. This is illustrated by Fig.~\ref{example}(a) which represents one of diagrams describing the effect of the electron-photon scattering on the electron current. The vertical broken line shows the cut to be made to relate this diagram with the cross-section of the Compton scattering (note that all photon momenta in this diagram are on the mass shell). This effect will be considered in detail in Sec.~\ref{relaxation}, where it will be shown explicitly that the expression for the electron-photon scattering cross-section is contained in diagrams similar to that in Fig.~\ref{example}(a), appearing in the kinetic equation for the electron momentum probability distribution. Analogously, shown in Fig.~\ref{example}(b) is one of diagrams describing the effect of photon emission by the electron, that accompanies the scattering of heat-bath photons by the electron, {\it etc.}

Because of finiteness of the time contour $C,$ energy is not conserved in the interaction vertices. As a result, the usual $\delta$-functions expressing energy conservation in the S-matrix theory become smeared, so that the vertex factor takes the form, in the Coulomb gauge,
\begin{eqnarray}\label{vert_C}
- {\rm i} e \gamma^\mu \overline{\Delta}_t(v)\,,
\quad \overline{\Delta}_t(v) = \frac{{\rm e}^{{\rm i}v^0 t} - {\rm e}^{{\rm i}v^0 t_0}}{{\rm i}v_0}(2\pi)^3\delta^{(3)}(\bm{v})\,,
\end{eqnarray}
where $v$ is the sum of outgoing 4-momenta, to be referred to in what follows as the {\it residual momentum}. In the Lorentz gauge, the non-invariant term in $L_l$ modifies the vertex to
\begin{eqnarray}\label{vert_L}
- {\rm i} e \gamma^\alpha g^\mu_\alpha(k)\overline{\Delta}_t(v)\,,
\quad g^\mu_\alpha(k)=\delta^\mu_\alpha+\frac{k_\alpha\left(k^\mu-\eta^\mu k_0\right)}{\bm{k}^2}\,.
\end{eqnarray}
It is to be noted that the gauge invariance of $F$ implies $\xi$-independence of $F^{\rm eff},$ for $g^\mu_\alpha(k)k_\mu \equiv 0,$ so that $d_{\mu\nu}$ in the Lorentz gauge can be replaced by $\eta_{\mu\nu}$. Moreover, regarding the analytic structure of Feynman diagrams, the Lorentz gauge is completely equivalent to the Coulomb gauge when $F$ does not involve $A_{\mu}.$ In that case, there are only internal photon lines each of which has its ends contracted with the tensor $g^\mu_\alpha(k)$:
$$g^\mu_\alpha(k)\eta_{\mu\nu}g^\nu_\beta(k).$$ A simple calculation shows that the latter is exactly the expression (\ref{coulombd}) for $d_{\alpha\beta}$ in the Coulomb gauge.

As a result of the smearing of the energy $\delta$-function, the effective electromagnetic field and the effective current become infrared finite, so that no special regularization such as the $\lambda$-regularization introduced in \cite{jpa} is needed in investigating the infrared effects.

\subsection{Pole prescriptions and multiloop Feynman integrals}\label{polespr}

\begin{figure}[h]
\begin{center}
     \includegraphics{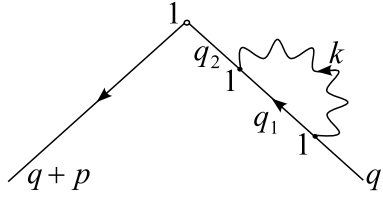}
\caption{A one-loop diagram contributing to the effective current.}
\label{selfa}
\end{center}
\end{figure}

Regarding pole prescriptions symbolized by $\pm{\rm i}0$ in Eqs.~(\ref{electronpr}), (\ref{photonpr}), it is worth to make the following technical remark. As we shall see below, evaluation of multiple Feynman integrals is often facilitated by the use of residual momenta as integration variables. However, this requires some care to avoid appearance of ambiguities when changing the order of integration. As an illustration, consider diagram in Fig.~\ref{selfa}. Its contribution is proportional to

\begin{eqnarray}\label{examp_order}
\int \frac{{\rm d}^4 k}{(2\pi)^4}\int\frac{{\rm d}^4 q_1}{(2\pi)^4}\int\frac{{\rm d}^4 q_2}{(2\pi)^4}
\frac{\overline{\Delta}_t(q_2-q_1-k)\overline{\Delta}_t(q_1-q+k)}{m-\slashed{q}_2-{\rm i} 0}\gamma^\alpha
\frac{D^{(11)}_{\alpha\beta}(k)}{m-\slashed{q}_1-{\rm i} 0}\gamma^\beta.
\end{eqnarray}\noindent The imaginary infinitesimals $(-{\rm i} 0)$ specify contours of integration over $q_1^0$ and $q_2^0,$ whose independence allows interchanging the order of integration. In order to go over to integration with respect to residual momenta $v_1,$ $v_2,$ we first change the integration variable $q_1 \to v_1$ according to $q_1=q-k+v_1,$ and then change $q_2 \to v_2$ using the relation $q_2=q+v_1+v_2,$ in which $v_1$ is treated as a complex parameter. Equation~(\ref{examp_order}) thus takes the form
\begin{eqnarray}
\int \frac{{\rm d}^4 k}{(2\pi)^4}\int\frac{{\rm d}^4 v_1}{(2\pi)^4}
\int_{v_1}\frac{{\rm d}^4 v_2}{(2\pi)^4}
\frac{\overline{\Delta}_t(v_1)\overline{\Delta}_t(v_2)}{m-\slashed{q}-\slashed{v}_1-\slashed{v}_2-{\rm i} 0}\gamma^\alpha
\frac{D^{(11)}_{\alpha\beta}(k)}{m-\slashed{q}+\slashed{k}-\slashed{v}_1-{\rm i} 0}\gamma^\beta\,,\nonumber
\end{eqnarray}\noindent where the subscript $v_1$ indicates that the position of integration contour over $v_2,$ specified by the left pole factor in the integrand, depends also on $v_1$ which runs the contour specified by the other pole factor. Interchanging the order of integration with respect to $v_1$ and $v_2$ now is not legitimate, as it leads to ambiguity in integrating the pole $1/[m^2-(q+v_1+v_2)^2]$ with respect to $q^0_1.$ But the last formula and its generalizations turn out to be useful even when they do not admit changing the order of integration.

We will have more to say on this issue in Secs.~\ref{approx}, \ref{factorization}.

\section{Infrared thermalization}\label{infrared_calc}

Upon transition to the momentum space, Eq.~(\ref{feff}) with $F=J_{\mu}$ takes the form
\begin{eqnarray}\label{JFur_e_1}
J^{\rm eff}_\mu(\bm{x},t) = \sum_{\sigma,\sigma'}\int
\frac{{\rm d}^3 \bm{q}}{(2\pi)^3} \frac{{\rm d}^3\bm{p}}{(2\pi)^3}\varrho_{\sigma\sigma'}(t;\bm{q},\bm{q}+\bm{p})
\frac{\bar{u}_{\sigma'}(\bm{q}+\bm{p})\gamma_\mu u_{\sigma}(\bm{q})}
{\sqrt{2\varepsilon_{\bm{q}+\bm{p}}}\sqrt{2\varepsilon_{\bm{q}}}}{\rm e}^{-{\rm i}\bm{p}\bm{x}}\,,
\end{eqnarray}\noindent
where $\varepsilon_{\bm{q}}=\sqrt{m^2+\bm{q}^2}$,
$p_0=\varepsilon_{\bm{q}+\bm{p}}-\varepsilon_{\bm{q}},$ and $\varrho(t;\bm{q},\bm{q}')$ is the electron density matrix in the momentum space representation; by definition,
\begin{eqnarray}\label{denmom_def}
\varrho_{\sigma\sigma'}(t;\bm{q},\bm{q}') = \langle e|\varrho(t)|e' \rangle,
\end{eqnarray}
where $\bm{q},\sigma$ are the electron momentum and spin in the state $|e\rangle,$ and $\bm{q}',\sigma'$ same for $|e'\rangle.$ This matrix is normalized on unity,
\begin{eqnarray}\label{normirovka}
\sum\limits_{\sigma}\int
\frac{{\rm d}^3 \bm{q}}{(2\pi)^3}\varrho_{\sigma \sigma}(\bm{q},\bm{q}) = 1\,.
\end{eqnarray}
Finally, the bispinor amplitudes $u_{\sigma}(\bm{q})$ are also normalized on unity, $\bar{u}_{\sigma}u_{\sigma}=1,$ and satisfy $$(\slashed{q} - m)u_{\sigma}(\bm{q}) = 0.$$

The effective electromagnetic field can be found from Eq.~(\ref{HeisAver}). Note that in the Coulomb gauge the scalar potential reads simply
\begin{eqnarray}
A^{\rm eff}_0(\bm{x},t) = e\sum_{\sigma,\sigma'}\int
\frac{{\rm d}^3 \bm{q}}{(2\pi)^3} \frac{{\rm d}^3\bm{p}}{(2\pi)^3}\varrho_{\sigma\sigma'}(t;\bm{q},\bm{q}+\bm{p})
\frac{\bar{u}_{\sigma'}(\bm{q}+\bm{p})\gamma_0 u_{\sigma}(\bm{q})}
{\sqrt{2\varepsilon_{\bm{q}+\bm{p}}}\sqrt{2\varepsilon_{\bm{q}}}}\frac{{\rm e}^{-{\rm i}\bm{p}\bm{x}}}{\bm{p}^2}\,.
\end{eqnarray}\noindent Using this expression and taking into account the gauge condition $\partial_i A_i=0,$ it is not difficult to verify that the effective electromagnetic field satisfies Gauss law in the infinite space.

\begin{figure}[h]
\begin{center}
     \includegraphics{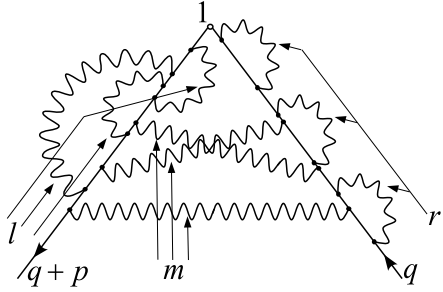}
\caption{General infrared-divergent contribution to the effective current. $q$ and $p$ are the electron 4-momentum and 4-momentum transfer, respectively.}
\label{sumj}
\end{center}
\end{figure}

\begin{figure}[h]
\begin{center}
     \includegraphics{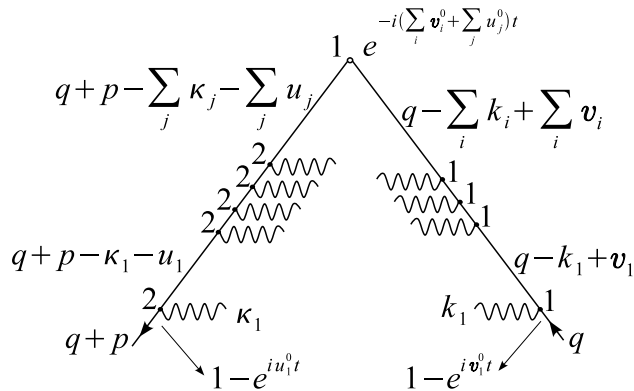}
\caption{Detailed structure of diagrams contributing to the effective current. The horizontal photon lines are supposed to be arbitrarily paired.}
\label{chan}
\end{center}
\end{figure}

The contributions we are interested in come from integration over small virtual and residual momenta. The results of \cite{jpa} imply that these contributions diverge in the limit $t_0\to-\infty,$ or equivalently $t\to\infty,$ as $\ln(mt)$ in the vacuum case, and as $Tt$ at finite temperature. Typical infrared-singular diagram is shown in Fig.~\ref{sumj}. Its internal structure is further specified by Fig.~\ref{chan}, where $k,$ $\kappa$ denote virtual momenta, and $v,$ $ u$ the residual momenta; henceforth, we set $t_0=0,$ without restricting generality. The factor
$\exp-{\rm i}(\sum_i v^0_i+\sum_j  u^0_j)t$ is brought in by the current operator in the interaction picture, whereas factors $(1-\exp{\rm i}v_i^0 t),$ $(1-\exp{\rm i} u_j^0 t)$ come from the vertex functions $\overline{\Delta}_t.$ It is easily verified that  the product of all these factors is invariant under the following replacement
\begin{eqnarray}
{\rm e}^{-{\rm i}(\sum_i v^0_i+\sum_j  u^0_j)t}\to1\,,
\quad \overline{\Delta}_t(v)\to
\Delta_t(v)={\rm i}\frac{{\rm e}^{-{\rm i}v_0 t}-1}{v_0}(2\pi)^3\delta^{(3)}(\bm{v})\,,
\end{eqnarray}\noindent a fact that will be used below.

\subsection{Approximations}\label{approx}

Since we are interested in the low-energy properties of the effective current, and since we have assumed nonrelativistic conditions for the electron, in what follows we shall systematically neglect radiative corrections to the electron and photon self-energy as well as to their interaction, but only if they represent {\it finite relative} corrections to the effective field, that is, if they give rise to factors $[1 + O(|\bm{q}|/m)]$ or $[1 + O(|\bm{p}|/m)]$ in the leading term. Also, we shall use the condition $T \ll m$ underlying our model to omit terms proportional to $T/m,$ unless such term is leading. In this approximation, the momentum space density matrices at the instants $t$ and $t_0$ are related by
\begin{eqnarray}\label{varrhoo_R}
\varrho_{\sigma\sigma'}(t;\bm{q},\bm{q}')=
\varrho_{\sigma\sigma'}(t_0;\bm{q},\bm{q}')R(t;\bm{q},\bm{q}'){\rm e}^{{\rm i}p^0t}\,,
\end{eqnarray}
\noindent where the scalar factor $R(t;\bm{q},\bm{q}')$ incorporates radiative corrections ($R = 1$ in the tree approximation). Indeed, the $\gamma$-matrix structure of a $N$-loop infrared-divergent diagram contributing to $J^{\rm eff}_{\mu}$ is (see Fig.~\ref{chan})
\begin{eqnarray}
\sum_{\sigma,\sigma'}\varrho_{\sigma\sigma'}(t_0;\bm{q},\bm{q}')\bar{u}_{\sigma'}(\bm{q}')
\gamma^{\beta_1}(\slashed{q}'-\slashed{\kappa}_1-\slashed{ u}_1+m) \cdots\gamma^{\beta_{f}}(\slashed{q}' -
\sum_j[\slashed{\kappa}_j+\slashed{ u}_j]+m)
\gamma_\mu \nonumber\\
\times(\slashed{q}-\slashed{k}_1+\slashed{v}_1+m) \gamma^{\alpha_1}\cdots(\slashed{q}-\sum_i[\slashed{k}_i-\slashed{v}_i]+m)\gamma^{\alpha_{s}}
u_{\sigma}(\bm{q}), \nonumber
\end{eqnarray}
where $s + f = 2N$. At zero temperature, the Feynman integrals diverge logarithmically, so that $\slashed{k},$ $\slashed{v}$, etc. in this expression give rise to infrared-finite terms. For $T\ne 0,$ the divergence is linear, and the corresponding subleading contribution diverges logarithmically. On dimensional grounds, its relative order is $O(T/m),$ so that this contribution can be omitted. Thus, the above expression can be replaced by
\begin{eqnarray}&&
\sum_{\sigma,\sigma'}\varrho_{\sigma\sigma'}(t_0;\bm{q},\bm{q}')\bar{u}_{\sigma'}(\bm{q}')
\gamma^{\beta_1}(\slashed{q}'+m) \cdots\gamma^{\beta_{f}}(\slashed{q}'+m)
\gamma_\mu(\slashed{q}+m) \gamma^{\alpha_1}\cdots(\slashed{q}+m)\gamma^{\alpha_{s}}
u_{\sigma}(\bm{q}) \nonumber \\&&
= \sum_{\sigma,\sigma'}\varrho_{\sigma\sigma'}(t_0;\bm{q},\bm{q}') 2q^{\prime\beta_1}\cdots 2q^{\prime\beta_{f}}2q^{\alpha_1}\cdots 2q^{\alpha_{s}}
\bar{u}_{\sigma'}(\bm{q}')\gamma_\mu u_{\sigma}(\bm{q}),\nonumber
\end{eqnarray}\noindent from which Eq.~(\ref{varrhoo_R}) follows. This relation will be derived in a more direct way in Sec.~\ref{effmat_calc}. The above reasoning also implies a similar simplification in denominators of the electron propagators
\begin{eqnarray}\nonumber
\frac{1}{m^2 - (q - \sum k_i + \sum v_i)^2 \pm {\rm i}0}\to
\frac{1}{2q(\sum k_i - \sum v_i) \pm {\rm i}0}, \quad {\rm etc.}
\end{eqnarray}\noindent Next, we note that graphs with a 1-vertex appearing to the left of a 2-vertex can be omitted, because they involve the function $D^{(12)}(q)\sim\theta(-q_0),$ and therefore do not contribute at small loop momenta. In particular, all vertices on the incoming electron line (the right slope of diagram in Fig.~\ref{chan}) must be type-1. Less trivial is the fact that all vertices on the outgoing electron line (the left slope of the diagram in Fig.~\ref{chan}) are to be of type-2 to give rise to a nonvanishing contribution. To see this, imagine for a moment that the rightmost vertex on the outgoing electron line is of type-1. Then following the recipe formulated in Sec.~\ref{polespr}, we first perform integration over $u_f,$ and find that
\begin{eqnarray}\nonumber
\int\frac{{\rm d} u^0_f}{2\pi{\rm i}}\frac{{\rm e}^{-{\rm i} u_f^0 t}-1}{ u_f^0}\frac{1}{2q(\sum\limits_j \kappa_j+\sum\limits_{j\ne f}  u_j +  u_f) - {\rm i}0} = 0,
\end{eqnarray}\noindent because the contour of integration can be closed in the lower half-plane of complex $ u_f^0$ (recall that $t>0$). Therefore, the rightmost vertex on the left slope must be type-2, which proves the assertion, for as was shown before, vertices to the left of a 2-vertex must be type-2.

\subsection{Factorization of infrared contributions}\label{factorization}

\begin{figure}[b]
     \includegraphics{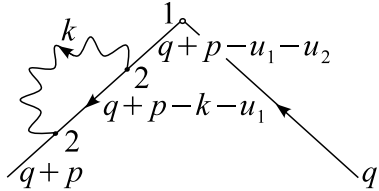}
\caption{The one-loop diagram determining the function $F^0_{21}.$}
\label{selfb}
\end{figure}

\begin{figure}[h]
\begin{center}
     \includegraphics{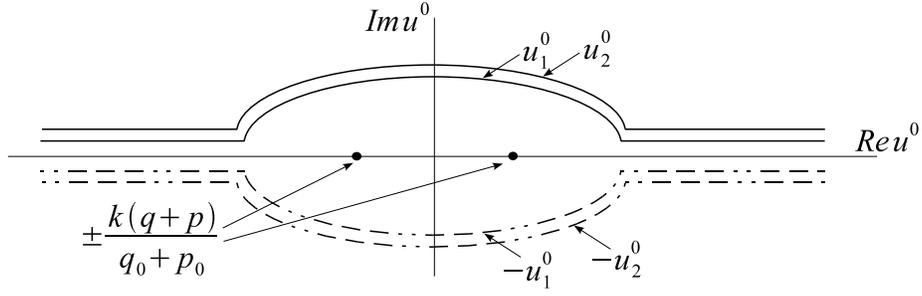}
\caption{Pole structure and contours of integration in the complex $u^0$-plane in the integral representing $F^0_{21}.$}
\label{poles}
\end{center}
\end{figure}

To separate contributions singular in the limit $t\to \infty,$ we use the condition $T\ll m$ to introduce a momentum threshold $\Lambda$ such that $T\ll \Lambda\ll m,$ which identifies the photons with $0 < k < \Lambda$ as ``soft.'' As was already mentioned in Sec.~\ref{rules}, our model requires no special infrared regularization, because restriction to a finite time interval renders all  Feynman integrals convergent at small momenta. As to ultraviolet divergences, they are supposed to be regularized using some conventional means, say, the Pauli-Villars  technique. If, as usual, the corresponding masses are chosen larger than the electron mass, then the ultraviolet divergences can be isolated and subtracted without affecting infrared properties of the theory, and we will assume that this has been done.
Using the momentum threshold, the perturbation series for the function $R(t;\bm{q},\bm{q}+\bm{p})$ can be written as
\begin{eqnarray}\label{R_I}
R(t;\bm{q},\bm{q}+\bm{p}) = I_{\Lambda}(\bm{p},\bm{q})\sum_{N=0}^{\infty}(e^2)^N I_N (\bm{p},\bm{q},\Lambda),
\end{eqnarray}\noindent
where the factor $I_{\Lambda}(\bm{p},\bm{q})$ is the contributions of photons with momenta $k > \Lambda,$ and $I_N(\bm{p},\bm{q},\Lambda)$ are the infrared-singular parts of diagrams of the type shown in Fig.~\ref{sumj}. Introducing abridged notation $Q_{st}(k) = q^{\mu}_s q^{\nu}_t D^{(st)}_{\mu\nu}(k),$ where $q_1=q, q_2=q+p,$ and $s,t$ take on values $1,2,$ the functions $I_N$ take the form
\begin{eqnarray}\label{I_N}
I_N (\bm{p},\bm{q},\Lambda)=
\frac{1}{{\rm i}^N}\sum_{r=0}^N &&\hspace{-0,3cm} \sum_{l=0}^{N-r}\int\limits^{\Lambda} \prod_{i=1}^{m}\frac{{\rm d}^4 k_i}{(2\pi)^4}Q_{21}(k_i) \frac{F_{1r}^{m}(k_1,\dots,k_{m})F_{2l}^{m}(k_1,\dots,k_{m})}{r!2^{r}l!2^{l}m!} \end{eqnarray}\noindent
where $N$ is the number of virtual photon lines, of which $r$ ($l$) reside on the incoming (outgoing) electron line, while the remaining $m \equiv N-r-l$ connect the two electron lines; the symbol $\int\limits^{\Lambda}$ indicates that all loop integrals are cut off at $k=\Lambda$; finally, the functions $F_{2l}^{m}$ and $F_{1r}^{m}$ incorporate electron propagators and vertex factors. They include sums over all permutations of vertices residing on the outgoing and incoming electron line, respectively, the factor $r!2^{r}l!2^{l}$ accounting for redundant permutations. To put them in a form admitting factorization of multiloop diagrams, we proceed as explained in Sec.~\ref{polespr} and go over to integration with respect to residual momenta. To be specific, consider the lowest order diagram shown in Fig.~\ref{selfb}. 
Writing
$$2\pi {\rm i}\delta(m^2-q^2) = \frac{1}{m^2-q^2-{\rm i}0} - \frac{1}{m^2-q^2+{\rm i}0}$$ for the factor brought in by $D^{(21)},$ we see that the first term does not contribute (Cf. discussion at the end of Sec.~\ref{approx}), whereas the second term gives
\begin{eqnarray}\label{poles1}
F^0_{21}=\int\limits^{\Lambda} \frac{{\rm d}^4 k}{(2\pi)^4}Q_{22}(k)
\int\limits_k\frac{{\rm d}^4 u_1}{(2\pi)^4}\frac{\Delta_t(u_1)}{(k+u_1)q_2 + {\rm i} 0}
\int\limits_{u_1}\frac{{\rm d}^4 u_2}{(2\pi)^4}\frac{\Delta_t(u_2)}{(u_1+u_2)q_2 + {\rm i} 0} + (u_1 \leftrightarrow u_2)\,, \nonumber
\end{eqnarray}\noindent where $(u_1 \leftrightarrow u_2)$ denotes the first term with $u_1,u_2$ interchanged. We observe that the contours of integration with respect to $u_1^0,$ $u_2^0$ can be chosen so as to meet the pole prescriptions in both terms simultaneously. Namely, the $u_1^0$-contour must go above the poles $\pm kq_2/q^0_2$, $-u_2^0,$ and not intersect the $u_2^0$-contour going above the poles $\pm kq_2/q^0_2$, $-u_1^0,$ as shown in Fig.~\ref{poles}. Taking into account that the function $Q_{22}(k)$ is even, and introducing new variables $w_1 = u_1 + k,$ $w_2 = u_2 - k$ yields

\begin{eqnarray}
F^0_{21} = \int\limits^{\Lambda}\frac{{\rm d}^4 k}{(2\pi)^4}&&\hspace{-0.3cm}
\int\frac{{\rm d}^4 u_1}{(2\pi)^4}\int\frac{{\rm d}^4 u_2}{(2\pi)^4}Q_{22}(k)
\frac{\Delta_t(u_1)\Delta_t(u_2)}{(u_1+u_2)q_2+{\rm i} 0}
\nonumber\\&&
\times\left[\frac{1}{(k+u_1)q_2 + {\rm i} 0}+\frac{1}{(-k+u_2)q_2 + {\rm i} 0}\right]
\nonumber\\
= \int\limits^{\Lambda}\frac{{\rm d}^4 k}{(2\pi)^4}&&\hspace{-0.3cm}
\int\frac{{\rm d}^4 w_1}{(2\pi)^4}\int\frac{{\rm d}^4 w_2}{(2\pi)^4}Q_{22}(k)
\frac{\Delta_t(w_1-k)}{w_1q_2 + {\rm i} 0}\frac{\Delta_t(w_2+k)}{w_2q_2 + {\rm i} 0}\,.
\nonumber
\end{eqnarray}\noindent This consideration is readily extended to all $l,m.$ The factor $F_{2l}^{m}$ can be written in general as $(w_1 +\cdots +w_k \equiv W_k)$
\begin{eqnarray}&&
\hspace{-1cm}F_{2l}^{m}(k_1,\dots,k_{m}) = \nonumber\\&& \hspace{-1cm}\int \frac{{\rm d}^4 w_1}{(2\pi)^4}\cdots \frac{{\rm d}^4 w_{2l+m}}{(2\pi)^4}\int\limits^{\Lambda}\prod_{i=1}^{l}\frac{{\rm d}^4 k_{m+i}}{(2\pi)^4}Q_{22}(k_{m+i})\Delta_t(w_{2i-1}-k_{m+i})\Delta_t(k_{m+i}+w_{2i})
\nonumber \\&& \hspace{-0.6cm}\times \Delta_t(w_{2l+1}-k_1)\cdots \Delta_t(w_{2l+m}-k_{m})
\sum\limits_{\rm perm}\left[ \frac{1}{W_1 q_2+{\rm i}0} \frac{1}{W_2 q_2+{\rm i}0}\cdots\frac{1}{W_{2l+m}q_2+{\rm i}0} \right],\nonumber
\end{eqnarray}\noindent where the sum is over all permutations of indices $1,2,\dots,2l+m.$ That this sum appears under the sign of integral is precisely because the imaginary infinitesimals in all its terms are of the same sign. Therefore, this sum can be factorized using the formula
\begin{eqnarray}\label{sumfactor}
\sum\limits_{\rm perm}\left[ \frac{1}{W_1 q +{\rm i}0} \frac{1}{W_2 q +{\rm i}0}\cdots\frac{1}{W_{m}q +{\rm i}0} \right] =
\frac{1}{w_1 q+{\rm i}0}\cdots\frac{1}{w_{m}q+{\rm i}0}\,,
\end{eqnarray}\noindent which is easily proved by induction for all integers $m.$
Similarly,
\begin{eqnarray}&&
\hspace{-1cm}F_{1r}^{m}(k_1,\dots,k_{m}) = \nonumber\\&& \hspace{-1cm}\int \frac{{\rm d}^4 w_1}{(2\pi)^4}\cdots \frac{{\rm d}^4 w_{2r+m}}{(2\pi)^4} \int\limits^{\Lambda}\prod_{i=1}^{r}\frac{{\rm d}^4 k_{m+i}}{(2\pi)^4}Q_{11}(k_{m+i})\Delta_t(k_{m+i}-w_{2i-1})\Delta_t(-k_{m+i}-w_{2i})
\nonumber \\&& \hspace{-0.6cm}\times \Delta_t(k_1-w_{2r+1})\cdots \Delta_t(k_{m}-w_{2r+m})
\sum\limits_{\rm perm}\left[ \frac{1}{W_1 q_1-{\rm i}0} \frac{1}{W_2 q_1-{\rm i}0}\cdots\frac{1}{W_{2r+m}q_1-{\rm i}0} \right],\nonumber
\end{eqnarray}\noindent where the sum is over all permutations of indices $1,2,\dots,2r+m,$ and factorization is accomplished using the complex conjugate of Eq.~(\ref{sumfactor}).
We thus obtain
\begin{eqnarray}
\hspace{-1cm}F_{2l}^{m}(k_1,\dots,k_{m}) =
\left[\int\limits^{\Lambda} \frac{{\rm d}^4 k}{(2\pi)^4}\int\frac{{\rm d}^4 w_1}{(2\pi)^4}\frac{{\rm d}^4 w_2}{(2\pi)^4}Q_{22}(k)\frac{\Delta_t(w_1-k)\Delta_t(w_2+k)}{(w_1q_2+{\rm i}0)(w_2q_2+{\rm i}0)}\right]^l \nonumber\\
\times\prod_{i=1}^{m}\int\frac{{\rm d}^4 w_{2n+i}}{(2\pi)^4}\frac{\Delta_t(w_{2l+i}-k_i)}{w_{2l+i}q_2+{\rm i}0}\,, \nonumber
\end{eqnarray}\noindent
\begin{eqnarray}\nonumber
\hspace{-1cm}F_{1r}^{m}(k_1,\dots,k_{m}) =
\left[\int\limits^{\Lambda} \frac{{\rm d}^4 k}{(2\pi)^4}\int\frac{{\rm d}^4 w_1}{(2\pi)^4}\frac{{\rm d}^4 w_2}{(2\pi)^4}Q_{11}(k)\frac{\Delta_t(-w_1+k)\Delta_t(-w_2-k)}{(w_1q_1-{\rm i}0)(w_2q_1-{\rm i}0)}\right]^r \nonumber\\
\times\prod_{i=1}^{m}\int\frac{{\rm d}^4 w_{2n+i}}{(2\pi)^4}\frac{\Delta_t(k_i-w_{2r+i})}{w_{2r+i}q_1-{\rm i}0}\,. \nonumber
\end{eqnarray}\noindent Finally, substitution into Eq.~(\ref{I_N}) gives
\begin{eqnarray}\nonumber
I_N (\bm{p},\bm{q},\Lambda)=\sum_{r=0}^N \sum_{l=0}^{N-r}
\frac{g_{11}^{r}}{r!\,2^{r}} \frac{g_{22}^{l}}{l!\,2^{l}} \frac{g_{21}^{m}}{m!} = \frac{g^N}{N!\,2^N},
\quad g \equiv g_{11}+g_{22}+2g_{21},
\end{eqnarray}\noindent
where
\begin{eqnarray}\nonumber
g_{rs} = -{\rm i}\eta_r \eta_s
\int\limits^{\Lambda} \frac{{\rm d}^4 k}{(2\pi)^4} \int\frac{{\rm d}^4 w_1}{(2\pi)^4}\frac{{\rm d}^4 w_2}{(2\pi)^4}Q_{rs}(k)\frac{\Delta_t(w_1-k) \Delta_t(w_2+k)}{(w_1 q_r+{\rm i}0)(w_2 q_s+{\rm i}0)}, \quad \eta_1=1, \eta_2=-1.
\end{eqnarray}\noindent
Thus,
\begin{eqnarray}\label{inpq}
R(t;\bm{q},\bm{q}+\bm{p}) = I_{\Lambda}(\bm{p},\bm{q})\exp{\frac{e^2g}{2}}\,.
\end{eqnarray}
Using the formula
\begin{eqnarray}\nonumber
\int\frac{{\rm d}^4 w}{(2\pi)^4}\frac{\Delta_t(w-k)}{wq+{\rm i}0}=\int\frac{{\rm d}w_0}{2\pi {\rm i}}\frac{e^{-{\rm i}(w_0-k_0)t}-1}{w_0-k_0}\frac{1}{w_0q^0-\bm{k q}+{\rm i}0}=\frac{{\rm e}^{{\rm i}t kq/q^{0}}-1}{kq}\,,
\end{eqnarray}\noindent
we find
\begin{eqnarray}\nonumber
g_{rs}={\rm i}\eta_r \eta_s \int\limits^{\Lambda} \frac{{\rm d}^4 k}{(2\pi)^4} \frac{Q_{rs}(k)}{(kq_r) (kq_s)}
\left[{\rm e}^{{\rm i}tkq_r/q^0_{r}}-1\right]\left[{\rm e}^{-{\rm i}tkq_s/q^0_{s}}-1\right].
\end{eqnarray}\noindent
Inserting explicit expressions (\ref{photonpr}) for the photon propagator, the last expression can be put in the form
\begin{eqnarray}
g_{rs}=\eta_r \eta_s \int\limits^{\Lambda} \frac{{\rm d}^4 k}{(2\pi)^3}\left[\theta(k_0)+n(\bm{k})\right]\delta(k^2)
\frac{d_{\mu\nu}(k)q_r^\mu q_s^\nu}{(kq_r) (kq_s)}
\left[1+{\rm e}^{{\rm i}t\left(kq_r/q^0_{r}-kq_s/q^0_{s}\right)}\right.\nonumber\\
\left.-{\rm e}^{-{\rm i}t\eta_r kq_r/q^0_{r}}-{\rm e}^{-{\rm i}t\eta_s kq_s/q^0_{s}}\right],\nonumber
\end{eqnarray}\noindent which can be checked by inspection. Writing $g_{rs}=g^0_{rs}+g^T_{rs},$ where $g^0_{rs}$ is the limit of $g_{rs}$ for $T\to 0$ and $t$ fixed, the result of evaluation of these integrals in various limiting cases reads (assuming $\bm{q}\ll m,$ $\bm{p}\ll m$)
\begin{eqnarray}\label{res1}
g^0=
\left\{
\begin{array}{lc}
\displaystyle 0,& mt\ll 1, \\
\displaystyle -\frac{1}{3\pi^2}\frac{\bm{p}^2}{m^2}\ln\Lambda t,&mt\gg 1,|\bm{p}|t\ll 1, \\
\displaystyle -\frac{1}{3\pi^2}\frac{\bm{p}^2+\bm{q}^2+\bm{p}\bm{q}}{m^2}\ln\Lambda t,&|\bm{p}|t\gg 1,
\end{array}
\right.
\end{eqnarray}\noindent

\begin{eqnarray}\label{res2}
g^T=
\left\{
\begin{array}{lc}
\displaystyle 0,& Tt\ll 1, \\
\displaystyle \frac{1}{3\pi^2}\frac{\bm{p}^2}{m^2}\ln Tt
-\frac{1}{3\pi}\frac{\bm{p}^2}{m^2}Tt,&Tt\gg 1,T{|\bm{p}|}t/{m}\ll 1, \\
\displaystyle \frac{1}{3\pi^2}\frac{\bm{p}^2+\bm{q}^2+\bm{p}\bm{q}}{m^2}\ln Tt
-\frac{1}{3\pi}\frac{\bm{p}^2}{m^2}Tt,&T{|\bm{p}|}t/{m}\gg 1.
\end{array}
\right.
\end{eqnarray}\noindent
Thus, we arrive at the following expression for the large-time asymptotic of the function $R(t;\bm{q},\bm{q}+\bm{p})$
\begin{eqnarray}\label{result}
R(t;\bm{q},\bm{q}+\bm{p}) = I(\bm{p},\bm{q},t)
\exp\left(-\frac{2\alpha\bm{p}^2}{3m^2}Tt\right), \quad \alpha = \frac{e^2}{4\pi}\,, \quad Tt\gg1.
\end{eqnarray}\noindent
where $I(\bm{p},\bm{q},t)$ is the $\Lambda$-independent product of $I_{\Lambda}(\bm{p},\bm{q})$ with the factor contributed by the term proportional to $\ln\Lambda t$ in Eq.~(\ref{res1}). Since $\Lambda\ll m,$ within the logarithmic accuracy this amounts to replacing $\ln\Lambda t \to \ln mt.$ Conditions $T\ll m$ and $Tt\gg 1$ then imply that the logarithmic term is negligible in comparison with the term proportional to $(Tt).$ Thus, as long as the first stage of the electron evolution is considered, one can set $I(\bm{p},\bm{q},t)=1.$

It follows from Eq.~(\ref{result}) that $R(t;\bm{q},\bm{q})$ is infrared-finite, which in view of Eq.~(\ref{varrhoo_R}) means that the infrared effects do not change the probability distribution of the electron momentum. The vanishing of the infrared-divergent contribution at $\bm{p}=0$ is actually a consequence of the total charge conservation, as can be seen from Eqs.~(\ref{JFur_e_1}), (\ref{varrhoo_R}). It is to be noted also that the infrared finiteness of the diagonal elements of the electron density matrix can be viewed as a special case of the Bloch-Nordsieck theorem [apply this theorem to Eq.~(\ref{diag})].

\section{Alternative derivation of Eq.~(\ref{result})}\label{effmat_calc}

The exponential time dependence of the main result expressed by Eq.~(\ref{result}) suggests that it can be derived from consideration of the density matrix evolution in infinitesimal form. We give this derivation below in order to emphasize the role of the approximations made, and to demonstrate a very important fact that the infrared properties of currents can be determined without having to sum the perturbation series explicitly.

\begin{figure}[b]
\begin{center}
     \includegraphics{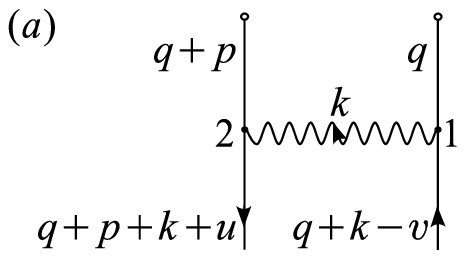}
     \includegraphics{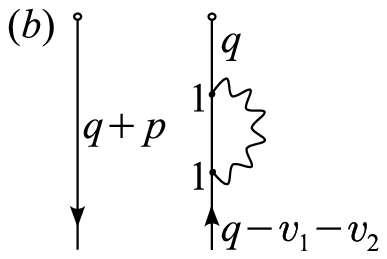}
     \includegraphics{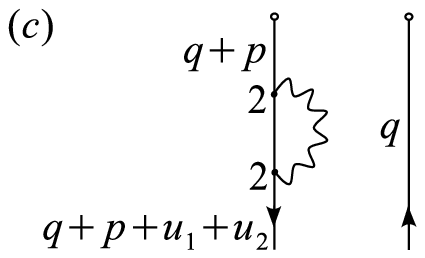}
\caption{Diagrams representing the second-order contribution to the integral kernel  $K(\bm{q},\bm{p},\bm{k})$ in Eq.~(\ref{Ro_K_diff}).}
\label{rho}
\end{center}
\end{figure}

Let us first establish a general relation between the matrices $\varrho(t)$ and $\varrho_0.$ It follows from Eq.~(\ref{densitymatrix}) that in the momentum representation, these matrices are related by
\begin{eqnarray}\label{densitymatrixm}
\varrho(t;\bm{q},\bm{q}') &=& \sum_{\phi,\phi_0}w(\phi_0) \langle  \phi|\langle e|U(t)|\psi_0\rangle|\phi_0\rangle\langle \phi_0|\langle \psi_0|U^\dag(t)|e' \rangle|\phi\rangle \nonumber \\
&=& \sum_{\phi,\phi_0,e_0,e'_0}w(\phi_0) \langle  \phi|\langle e|U(t)|e_0\rangle|\phi_0\rangle\langle e_0|\varrho_0| e'_0 \rangle\langle\phi_0|\langle e'_0|U^\dag(t)|e' \rangle|\phi\rangle,
\end{eqnarray}\noindent where $U(t)\equiv U(t,0),$ and
$$\langle e_0|\varrho_0| e'_0 \rangle = \langle e_0|\psi_0\rangle \langle \psi_0|e'_0 \rangle = \varrho(0;\bm{q}_0,\bm{q}'_0).$$ For simplicity, the electron is assumed henceforth unpolarized, $\varrho_{\sigma\sigma'} \sim \delta_{\sigma\sigma'},$ and the spin indices are omitted. By virtue of momentum conservation, the evolution operator has nonvanishing matrix elements only between states satisfying $$\bm{q}_0' = \bm{q}' + \bm{Q}, \quad \bm{q}_0 = \bm{q} + \bm{Q}$$
where $\bm{Q}$ denotes the total momentum of photons in a state $|\phi\rangle,$ and it is taken into account that this momentum is zero initially, $\bm{Q}_0 = 0.$ It follows that
$$\bm{q}_0' - \bm{q}_0 = \bm{q}'- \bm{q} \equiv \bm{p}.$$
Hence, the momentum space density matrices at the instants $t$ an $t_0=0$ are related by
\begin{eqnarray}\label{ro_omega}
\varrho(t;\bm{q},\bm{q}+\bm{p})=
\int\frac{{\rm d}^3 \bm{k}}{(2\pi)^3}
K(t;\bm{q},\bm{p},\bm{k})
\varrho(0;\bm{q} + \bm{k},\bm{q}+\bm{p} + \bm{k}).
\end{eqnarray}\noindent
The kernel $K(t;\bm{q},\bm{p},\bm{k})$ can be computed using the general formulas given in Secs.~\ref{generalformulation}, \ref{rules}, for which purpose it is useful to rewrite definition (\ref{denmom_def}) as
\begin{eqnarray}\label{vrrho_useful}
\varrho(t;\bm{q},\bm{q}')=
\int{\rm d}^3\bm{x}\int{\rm d}^3\bm{x}'
\sqrt{2\varepsilon_{\bm{q}'}}\sqrt{2\varepsilon_{\bm{q}}}
{\rm e}^{{\rm i}\bm{q}'\bm{x}'-{\rm i}\bm{q}\bm{x}}
\langle\bar{\psi}(t,\bm{x}')u(\bm{q}')\bar{u}(\bm{q})\psi(t,\bm{x})\rangle.
\end{eqnarray}\noindent
To determine the large-time behavior of $\varrho(t),$ one may proceed as in Sec.~\ref{factorization}, prove factorization of the infrared contributions, and then sum the perturbation series. However, the leading term of the large-time asymptotic can be found more directly. As was shown in Sec.~\ref{factorization}, this term is produced by interaction of the electron with equilibrium photons. To extract this term, we note that the property of being in equilibrium implies that the effect of such photons is homogeneous in time, but with one important qualification. The photon cloud surrounding the electron can be said to be in equilibrium at the given temperature $T$ only when considered on time intervals $\gg 1/T.$ This is because at lesser time intervals ($\lesssim 1/T$), the quantum indeterminacy in the photon energy becomes of order of the photon mean energy, and in the presence of such large fluctuations it is evidently impossible to speak about time homogeneity. Thus, we introduce a time $\tau_0$ satisfying
\begin{eqnarray}\label{cond1}
T\tau_0 \gg 1,
\end{eqnarray}\noindent and restrict consideration to time intervals $\delta t \gtrsim \tau_0.$ This condition justifies omission of the logarithmic contributions due to vacuum photon-electron interaction, making thereby the photon impact on the electron evolution homogeneous in time. This implies that the electron density matrices at arbitrary instants $t$ and $t+\delta t$ are related by the same kernel as in Eq.~(\ref{ro_omega}), {\it viz.,}
\begin{eqnarray}\label{rhodeltat}
\varrho(t+\delta t;\bm{q},\bm{q}+\bm{p})=
\int\frac{{\rm d}^3 \bm{k}}{(2\pi)^3}
K(\delta t;\bm{q},\bm{p},\bm{k})
\varrho(t;\bm{q} + \bm{k},\bm{q}+\bm{p} + \bm{k}),
\end{eqnarray}\noindent
where $K$ is independent of $t.$ Whether or not this equation can be reduced to a differential equation depends on the structure of the kernel $K(\delta t;\bm{q},\bm{p},\bm{k}).$ The point is that in general, one cannot differentiate it with respect to $\delta t$ at $\delta t=0,$ because this would violate condition (\ref{cond1}). To put it differently, if both sides of this equation are expanded in powers of $\delta t,$ the question is whether the terms $O(\delta t^2)$ can be neglected in comparison with the linear term. That this is legitimate in the case under consideration is suggested by the results of the preceding section, and is confirmed by the subsequent computation. Thus, we expand $K(\delta t;\bm{q},\bm{p},\bm{k})$ up to the first order in $\delta t,$ and obtain
\begin{eqnarray}\label{Ro_K_diff}
\frac{\partial\varrho(t;\bm{q},\bm{q}+\bm{p})}{\partial t}=\int\frac{{\rm d}^3 \bm{k}}{(2\pi)^3}
K(\bm{q},\bm{p},\bm{k})\varrho(t;\bm{q} + \bm{k},\bm{q}+\bm{p} + \bm{k}),
\end{eqnarray}\noindent where
\begin{eqnarray}\label{K_deriv}
K(\bm{q},\bm{p},\bm{k})\equiv \left.\frac{\partial}{\partial t} K(t;\bm{q},\bm{p},\bm{k})\right|_{t=0}.
\end{eqnarray}\noindent In view of the smallness of the coupling constant, $e^2 \ll 1,$ the function $K(\bm{q},\bm{p},\bm{k})$ can be found in the second order approximation. Indeed, if we use Eq.~(\ref{vrrho_useful}) to relate $\varrho(t+\delta t)$ and $\varrho(t),$ where $\delta t \sim \tau_0,$ then the loop expansion of the integral kernel $K(\delta t;\bm{q},\bm{p},k)$ is a power series in $e^2 T \tau_0,$ since each loop divergences linearly, with the dominant contribution coming from the integration over photon momenta $k\sim 1/\tau_0.$ Therefore, contributions of higher order in $e$ are negligible, provided that $\tau_0$ satisfies
\begin{eqnarray}\label{cond2}
e^2 T\tau_0 \ll 1.
\end{eqnarray}\noindent It is because of the smallness of the coupling constant that  this condition is consistent with Eq.~(\ref{cond1}). In zeroth order in the coupling (free electron evolution), $\varrho(t;\bm{q},\bm{q}+\bm{p})\sim e^{{\rm i}p^0 t},$ implying that
$$\left.K(\bm{q},\bm{p},\bm{k})\right|_{e=0} = {\rm i} p^0(2\pi)^3\delta^{(3)}(\bm{k}).$$
Diagrams representing the second-order term in $K(\bm{q},\bm{p},\bm{k})$ are shown in Fig.~\ref{rho}. Using the definition (\ref{K_deriv}), this term can be written as
\begin{eqnarray}
\left.K(\bm{q},\bm{p},\bm{k})\right|_{e^2} =
-2\frac{e^2}{\tau_0} \int\frac{{\rm d} k^0}{2\pi}
n(\bm{k})\delta(k^2)\frac{d_{\mu\nu}(k)q_2^\mu q_1^\nu}{(kq_2) (kq_1)}
\left[{\rm e}^{{\rm i}\tau_0 kq_2/q^0_2}-1\right]
\left[{\rm e}^{-{\rm i}\tau_0 kq_1/q^0_1}-1\right]\nonumber\\
+\frac{e^2}{\tau_0} \delta^{(3)}(\bm{k})\int\frac{{\rm d}^4 k'}{2\pi}
n(\bm{k'})\delta(k'^2)\sum_{r=1,2}\frac{d_{\mu\nu}(k')q_r^\mu q_r^\nu}{(k'q_r) (k'q_r)}
\left[{\rm e}^{{\rm i}\tau_0 k'q_r/q^0_r}-1\right]
\left[{\rm e}^{-{\rm i}\tau_0 k'q_r/q^0_r}-1\right].
\nonumber
\end{eqnarray}\noindent

The leading large-time contribution in this integral comes from $\bm{k}\sim 1/\tau_0,$ and counting powers of $\bm{k}$ in Eq.~(\ref{Ro_K_diff}) readily shows that this contribution is independent of $\tau_0,$ as expected. It is to be emphasized in this connection that the four-dimensionality of spacetime is essential to reach this conclusion. In a three-dimensional spacetime, for instance, the leading term in the function $K(\delta t;\bm{q},\bm{p},\bm{k})$ is quadratic in $\delta t,$ and transition from Eq.~(\ref{rhodeltat}) to Eq.~(\ref{Ro_K_diff}) is not legitimate. We note, finally, that under condition (\ref{cond1}), $\bm{k}\sim 1/\tau_0 \ll T$ can be neglected in the arguments of $\varrho$ in Eq.~(\ref{Ro_K_diff}), so that in effect, $K(\bm{q},\bm{p},\bm{k})\sim \delta^{(3)}(\bm{k}).$ A computation similar to that performed at the end of Sec.~\ref{factorization} then gives
\begin{eqnarray}\nonumber
\left.K(\bm{q},\bm{p},\bm{k})\right|_{e^2} =
-\frac{e^2}{6\pi}\frac{T}{m^2}\bm{p}^2 (2\pi)^3\delta^{(3)}(\bm{k}).
\end{eqnarray}\noindent
Substituting this into Eq.~(\ref{Ro_K_diff}), we find
\begin{eqnarray}\nonumber
\frac{\partial}{\partial t}\varrho(t;\bm{q},\bm{q}+\bm{p})=
\left[{\rm i}p_0-\frac{e^2}{6\pi}\frac{T}{m^2}\bm{p}^2\right] \varrho(t;\bm{q},\bm{q}+\bm{p})\,,
\end{eqnarray}
and therefore,
\begin{eqnarray}\label{main_result}
\varrho(t;\bm{q},\bm{q}+\bm{p})=
\exp\left(-\frac{e^2}{6\pi}\frac{\bm{p}^2 T}{m^2}t\right){\rm e}^{{\rm i}p_0t}
\varrho(0;\bm{q},\bm{q}+\bm{p}),
\end{eqnarray}\noindent
in agreement with Eq.~(\ref{result}).

\section{Physical manifestations of the infrared singularity}\label{phexamples}

In this section, we discuss some physical applications of the main result expressed by Eq.~(\ref{main_result}).

\subsection{Evolution of a Gaussian wave-packet}\label{gausswave}

To illustrate the role of the irreversible charge spreading, consider an unpolarized electron prepared at $t_0=0$ in a pure state described in the momentum space by a Gaussian wavefunction
$$\phi_0(\bm{q}) = \left(4\pi l^2\right)^{3/4}\exp\{-l^2\bm{q}^2/2\}\,, \quad l = {\rm const}.$$ This wavefunction represents the electron localized in a region of characteristic size $l$ near the origin. The corresponding density matrix is
$\varrho_0(\bm{q},\bm{q}')=\tilde{\phi}_0(\bm{q}')\phi_0(\bm{q}).$ Substitution of this expression in Eqs.~(\ref{main_result}), (\ref{JFur_e_1}), and evaluation of the Gaussian integrals yields the following expression for the electron density at arbitrary instant $t$
\begin{eqnarray}\label{Theta}
J^{\rm eff}_{0}(\bm{x},t) = \frac{1}{\pi^{3/2}l^3_t}\exp\left\{-\frac{\bm{x}^2}{l^2_{t}}\right\}\,,
\quad l_{t}=\left(l^2+\frac{t^2}{m^2 l^2}+4\Theta t\right)^{1/2}\,,\quad
\Theta = \frac{2\alpha T}{3m^2}
\end{eqnarray}\noindent This result shows that in addition to the usual quantum spreading described by the term $t^2/m^2 l^2$ in $l_t,$ there is a spreading due to interaction of the electron with the heat-bath photons. It is easy to see that in the setting considered, the latter effect is dominated by the usual quantum-mechanical spreading. Indeed, for a given $t,$ the minimum of the sum $(l^2+t^2/m^2 l^2)$ is $2t/m,$ which is very large compared to $4\Theta\tau \sim \alpha T t/m^2,$ since $T\ll m,$ $\alpha \ll 1.$ But it is not difficult to give an example where the relation between the two effects is opposite. Namely, let the electron be prepared at $t_0 = 0$ in a pure state which is supposed to describe this electron localized at a later instant $t = \tau$ in a region of the size $l$ near a point ${\bm x}_0.$ Then an appropriate momentum-space amplitude in nonrelativistic quantum mechanics would be
$$\phi(\bm{q}) = \phi_0(\bm{q}){\rm e}^{-{\rm i}\bm{q}\bm{x}_0 + {\rm i}\varepsilon_{\bm{q}}\tau}\,,$$ where $\phi_0(\bm{q})$ is real; the factor ${\rm e}^{{\rm i}q_0\tau}$ realizes the free electron evolution backward in time from $t=\tau$ to $t=0.$ Taking $\phi_0$ as before, the effective electron density at the instant $t=\tau$ now is
$$J^{\rm eff}_{0}(\bm{x},\tau) = \frac{\displaystyle \exp\left\{-\frac{\bm{x}^2}{(l^2 + 4\Theta\tau)}\right\}}{\pi^{3/2}\left(l^2 + 4\Theta\tau\right)^{3/2}}\,,$$ which demonstrates that the actual size of the wave packet at $t=\tau$ is $l_{\tau} = (l^2+4\Theta\tau)^{1/2}.$ This simple example shows that the minimal uncertainty in the position of an electron evolving freely for a time $\tau$ is $\sim \sqrt{\Theta\tau}.$ Incidentally, this fact justifies the term ``irreversible spreading.'' It also implies that in the conventional approach based on the notion of infinitely remote past, the uncertainty is formally infinite, and the effective electromagnetic field is zero at any given spatial point.

We note also that the minimal uncertainty is proportional to the square root of the travel time, which is characteristic of diffusive processes. This suggests that the irreversible spreading of the electron wave-packet can be viewed as a kind of Brownian motion of the electron. However, this analogy cannot be taken literally, because the irreversible spreading is not driven by the electron-photon collisions. This is evident from the fact that it does not lead to relaxation of the electron momentum -- as is seen from Eq.~(\ref{main_result}), the momentum probability distribution $\varrho(t;\bm{q},\bm{q})$ is unaffected by the infrared singularity. Furthermore, the irreversibility of this effect shows itself also as an increase of quantum entropy of the electron state \cite{jpa}, which is directly related to the fact that the infrared singularity damps the off-diagonal elements of the electron density matrix. But in contrast to the ordinary statistics, this infrared thermalization takes place both at $T\ne 0$ and $T=0,$ though in vacuum the process is much slower (time dependence of the effective electron current is exponential at finite temperature, whereas in vacuum it is a power law). These circumstances emphasize the specifically quantum nature of the irreversible spreading, which has no proper analog in nonrelativistic physics.

How the usual relaxation of the electron momentum due to its collisions with the heat-bath photons is described in the present formalism will be shown in Sec.~\ref{relaxation}.

Conditions considered in the last example can be experimentally realized using a magnetic lens to focus an electron beam, but it is more advantageous to detect the effect of infrared thermalization as a decoherence of the electron waves in the classic two-slit experiment considered in the next section.

\subsection{Electron diffraction}\label{diffraction}
\begin{figure}[b]
\begin{center}
     \includegraphics{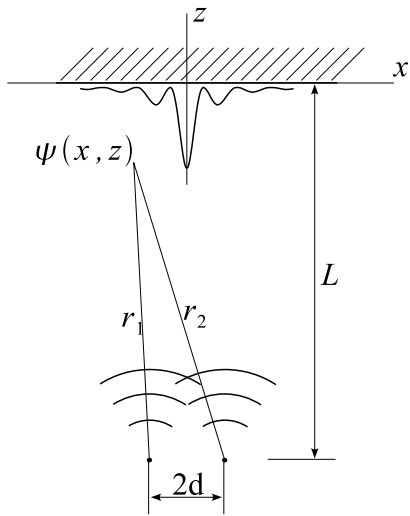}
\caption{Schematics of the two-slit experiment. The incident plane electron wave propagates upward; the slits are symbolized by small circles producing secondary electron waves.}
\label{exper}
\end{center}
\end{figure}
Consider the electron diffraction in the two-slit experiment shown schematically in Fig.~\ref{exper}. The wavefunction of diffracted electrons is a superposition of two outgoing cylindric waves, which at sufficiently large distance from the slits has the form
\begin{eqnarray}\nonumber
\psi=\frac{A}{\sqrt{2}}\left[\frac{{\rm e}^{{\rm i}kr_1}}{\sqrt{r_1}}+
\frac{{\rm e}^{{\rm i}kr_2}}{\sqrt{r_2}}\right],
\end{eqnarray}\noindent where $r_1$ ($r_2$) is the distance between the first (second) slit and the point of observation, $A$ is a bispinor amplitude independent of the spatial coordinates $x,z$ ($z$ being directed along the incident beam), and $k=|\bm{k}|$ is the incident electron momentum. Let $2d$ denote the slit spacing, and $L\gg d$ the distance between a slit and the screen at which we observe the interference pattern. In a vicinity of the detector ($x,z\ll L,$ assuming that the origin of the coordinate system is at the screen), the wavefunction can be written as
\begin{eqnarray}\nonumber
\psi(x,z)=\frac{\psi_0}{\sqrt{2}}{\rm e}^{{\rm i}kz}
\left[{\rm e}^{{\rm i}{\varkappa(x+d)^2}/4d}+{\rm e}^{{\rm i}{\varkappa(x-d)^2}/4d}\right]\,,
\quad \psi_0=\frac{A}{L}{\rm e}^{{\rm i}kL}\,, \quad \varkappa=\frac{2kd}{L}\,.
\end{eqnarray}\noindent Then in the absence of the photon bath, the electron density is given by
\begin{eqnarray}\nonumber
J_0(x)=\psi_0^\dag\psi_0\left[1+\cos(\varkappa x)\right]\,.
\end{eqnarray}\noindent To determine how this expression changes in the presence of thermal photons, we have to Fourier-expand the wavefunction
\begin{eqnarray}
\psi(x,z)=\sqrt{\frac{2\pi{\rm i}d}{\varkappa}}\psi_0{\rm e}^{{\rm i}kz}
\int\limits_{-\infty}^{+\infty}\frac{{\rm d}q}{2\pi}{\rm e}^{-{\rm i}d q^2/\varkappa}
\left[{\rm e}^{{\rm i}q(x+d)}+{\rm e}^{{\rm i}q(x-d)}\right].\nonumber
\end{eqnarray}\noindent  The corresponding expression for the electron density reads
\begin{eqnarray}
J_0(x)=\frac{8\pi d}{\varkappa}\psi_0^\dag\psi_0
\iint\limits_{-\infty}^{+\infty}\frac{{\rm d}q {\rm d}q'}{(2\pi)^2}
\exp{\rm i}\left[d(q'^2-q^2)/\varkappa-(q'-q)x\right]\cos (q'd) \cos (q d)\,.\nonumber
\end{eqnarray}\noindent Now, inclusion of the infrared effect of thermal photons gives for the electron density, according to Eqs.~(\ref{JFur_e_1}), (\ref{main_result}),
\begin{eqnarray}&&\nonumber
J_0^{\rm eff}(x)\nonumber\\&& = \frac{8\pi d}{\varkappa}\psi_0^\dag\psi_0
\iint\limits_{-\infty}^{+\infty}\frac{{\rm d}q {\rm d}q'}{(2\pi)^2}
\exp\left[-\Theta\tau(q'-q)^2+{\rm i}d(q'^2-q^2)/\varkappa-{\rm i}(q'-q)x\right]\cos(q'd) \cos(q d)\,,\nonumber
\end{eqnarray}
where $\Theta$ is defined in Eq.~(\ref{Theta}), and $\tau=mL/k$ is the electron travel time between the slits and the detector. Integrating back over $q,q',$ we find
\begin{eqnarray}\nonumber
J_0^{\rm eff}(x)=\psi_0^\dag\psi_0
\left[1+\exp\left(-\frac{2\alpha T L}{3m k}\varkappa^2\right)\cos(\varkappa x)\right]\,.
\end{eqnarray}\noindent The exponential factor in this formula describes the decoherence caused by the infrared electron thermalization. To determine conditions under which this effect is appreciable, we note that $\varkappa\sim 1/r,$ where $r$ is the spacing of the interference pattern. The interference is destroyed when the expression in the exponent becomes of order unity, or
$$\frac{T L}{\sqrt{\varepsilon}\,r^2}\sim 10^{20}\,{\rm K}/{\rm cm}\cdot{\rm eV}^{1/2}\,,$$
where $T$ is to be expressed in kelvins, and the electron energy $\varepsilon$ in electronvolts. The resolution of modern electron detectors employing magnetic lenses is a few angstrom. Therefore, for electron energy $\sim 10$\,eV and $L \sim 1$\,m, the effect is detectable already at $T\sim 100$\,K.

\section{Relaxation of the electron momentum}\label{relaxation}
\begin{figure}[h]
\begin{center}
     \includegraphics{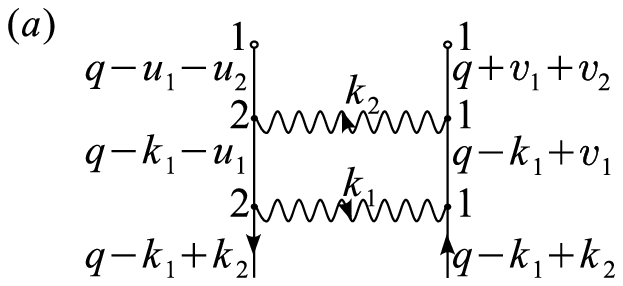}
     \includegraphics{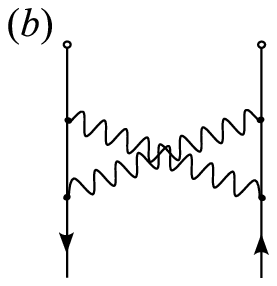}
     \includegraphics{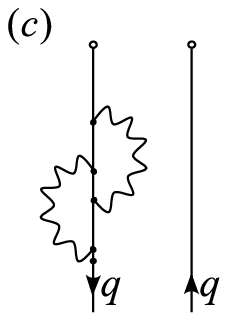}
     \includegraphics{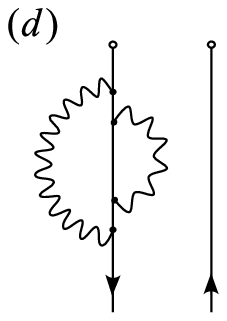}
\caption{Diagrams describing the effect of electron-photon scattering on the electron evolution. Open circles symbolize $\psi$-operators in the expression (\ref{vrrho_useful}) for the electron density matrix.}
\label{heat}
\end{center}
\end{figure}

To clarify the role played by the infrared thermalization, and to better expose its distinction from the usual thermalization, we shall now show how interaction of the electron with non-infrared photons realizes relaxation in the system, {\it i.e.,} how the electron momentum distribution tends to equilibrium distribution. To this end, we have to consider evolution of the diagonal elements of the electron density matrix, $\varrho(t;\bm{q},\bm{q}),$ which is described in the lowest order by diagrams shown in Fig.~\ref{heat}. As in Sec.~\ref{effmat_calc}, the leading contribution we are interested in turns out to be linear in time, so that $\partial\varrho(t;\bm{q},\bm{q})/\partial t$ can be written as a functional of $\varrho(t;\bm{q},\bm{q})$ at the same instant, as in Eq.~(\ref{Ro_K_diff}), with an integral kernel given exactly by diagrams in Fig.~\ref{heat}. A slight change of notation in this figure is to be noted, namely, the electron momentum $q$ is now off the mass shell, as it is associated with the $\psi$-operators in Eq.~(\ref{vrrho_useful}), symbolized in Fig.~\ref{heat} by open circles. On the contrary, the momentum $(q-k_1+k_2)$ is on the mass shell, as being associated with the external lines representing the initial density matrix. To express this fact, we write $q = \{q^0(k_1,k_2),\bm{q}\}\,,$ where
$q^0(k_1,k_2)=\varepsilon_{\bm{q}-\bm{k}_1+\bm{k}_2}+k_1^0-k_2^0.$ As we shall see, the leading contribution comes from integration over finite $k_1,$ $k_2,$ such that momentum $q$ is near the mass-shell. The two conditions $(q-k_1+k_2)^2=m^2$ and $q^2 = m^2$ are clearly consistent, since the equation $\varepsilon_{\bm{q}-\bm{k}_1+\bm{k}_2}=\varepsilon_{\bm{q}}-k_1^0+k_2^0$ has a continuum of non-trivial solutions with respect to $k_1 \ne k_2$ satisfying $k^2_1=k^2_2=0.$ The latter requirement follows from the fact that all vertices on the outgoing (incoming) electron line are of type 2 (1), by the same reason as in Sec.~\ref{approx}. It is to be noted also that $k_1^0$ and $k_2^0$ are of the same sign, as the opposite would allow for a double-photon emission by a free electron (formally, the relations $\varepsilon_{\bm{q}-\bm{k}_1+\bm{k}_2}=\varepsilon_{\bm{q}}-k_1^0+k_2^0,$ $q^2=m^2,$ $k^2_1=k^2_2 = 0$ are inconsistent for $k^0_1k^0_2<0$). Hence, no singularity of the type considered presently arise at zero temperature, as the vacuum contribution is proportional to  $\theta(k_1^0)\theta(-k_2^0).$

Evidently, contributions given by the disconnected diagrams in Figs.~\ref{heat}(c),\ref{heat}(d) do not change electron momentum, so that Eq.~(\ref{Ro_K_diff}) can be written as
\begin{eqnarray}\label{kineticeq}
\frac{\partial\varrho(t,\bm{q})}{\partial t}=
\int\frac{{\rm d}^3 \bm{k}}{(2\pi)^3}C(\bm{q},\bm{k})\varrho(t,\bm{q}+\bm{k})
+D(\bm{q})\varrho(t,\bm{q})\,,
\end{eqnarray}
where $\varrho(t,\bm{q})\equiv\varrho(t;\bm{q},\bm{q}),$ $C(\bm{q},\bm{k})$ is the connected part of the integral kernel, represented by diagrams in Figs.~\ref{heat}(a),\ref{heat}(b), and $D(\bm{q})$ is the disconnected part. The latter need not be calculated explicitly, since there is a simple relation between the two parts, following from the normalization condition (\ref{normirovka}). Namely, in order for this condition be satisfied by all solutions of Eq.~(\ref{kineticeq}), it is necessary that
\begin{eqnarray}\nonumber
D(\bm{q})= - \int\frac{{\rm d}^3 \bm{k}}{(2\pi)^3}C(\bm{q}+\bm{k},-\bm{k}),
\end{eqnarray}
and therefore,
\begin{eqnarray}\label{dRo_dt}
\frac{\partial\varrho(t,\bm{q})}{\partial t}=
\int\frac{{\rm d}^3 \bm{k}}{(2\pi)^3}
\left[C(\bm{q},\bm{k})\varrho(t,\bm{q}+\bm{k})-C(\bm{q}+\bm{k},-\bm{k})\varrho(t,\bm{q})\right]\,.
\end{eqnarray}\noindent This is nothing but the usual kinetic equation for an electron in the photon bath, with $C(\bm{q},\bm{k})$ playing the role of the probability of transition (per unit time) in which the electron goes from a state with momentum $(\bm{q}+\bm{k})$ over to a state with momentum $\bm{q}.$ To determine this function, we have to evaluate diagrams in Figs.~\ref{heat}(a),\ref{heat}(b). It is easy to see that the momenta $q-k_1,$ $q+k_2$ are off the mass shell whenever $q-k_1+k_2$ is on, and the light-like vectors $k_1,$ $k_2$ are nonzero. Therefore, the residual momenta can be neglected in the internal electron propagators when extracting the leading contribution:
$$\frac{\slashed{q}-\slashed{k}_1+\slashed{v}_1+m}{m^2-\left(q-k_1+v_1\right)^2+{\rm i}0}
\to\frac{\slashed{q}-\slashed{k}_1+m}{m^2-\left(q-k_1\right)^2}\,,\quad {\rm etc.}$$
Next, integration with respect to $k_1^0,$ $k_2^0$ yields four terms: each diagram contributes two terms -- one with $k^0_1 = +|\bm{k}_1|,$ $k^0_2 = +|\bm{k}_2|,$ and the other with $k^0_1 = -|\bm{k}_1|,$ $k^0_2 = -|\bm{k}_2|.$ Changing $\bm{k}_{1,2} \to - \bm{k}_{1,2}$ in the latter case, the connected contribution to the variation of the density matrix takes the form
\begin{eqnarray}
\delta_C\varrho(t,\bm{q})=
\int\frac{{\rm d}^3 \bm{k}_1}{(2\pi)^3}\int\frac{{\rm d}^3 \bm{k}_2}{(2\pi)^3}
\int\frac{\left[{\rm d} u\right]_{\delta t}\left[{\rm d} v\right]_{\delta t}
(2/\pi)\varepsilon^2_{\bm{q}}w(\bm{q},\bm{k}_1,\bm{k}_2)n(\bm{k}_1)[1+n(\bm{k}_2)]
\varrho(t,\bm{q}+\bm{k})}{[m^2-\left(q+v_1+v_2\right)^2-{\rm i}0][m^2-\left(q-u_1-u_2\right)^2+{\rm i}0]}\,,\nonumber\\ \label{dRo1}
\end{eqnarray}\noindent where
\begin{eqnarray}\label{w}
w(\bm{q},\bm{k}_1,\bm{k}_2)=&&\hspace{-0,3cm}
\frac{\pi e^4 d_{\mu\alpha}(k_1)d_{\nu\beta}(k_2)}
{2\varepsilon_{\bm{q}}2\varepsilon_{\bm{q}+\bm{k}}2|\bm{k}_1|2|\bm{k}_2|}\nonumber\\&&
\times{\rm tr}\left\{(\slashed{q}+\slashed{k}+m)
\left[\gamma^{\beta}\frac{\slashed{q}-\slashed{k}_1+m}
{m^2-\left(q-k_1\right)^2}\gamma^{\alpha}
+\gamma^{\alpha}\frac{\slashed{q}+\slashed{k}_2+m}
{m^2-\left(q+k_2\right)^2}\gamma^{\beta}\right]\right.\nonumber\\&&
\left.\hspace{1cm}\times(\slashed{q}+m)
\left[\gamma^{\mu}\frac{\slashed{q}-\slashed{k}_1+m}
{m^2-\left(q-k_1\right)^2}\gamma^{\nu}
+\gamma^{\nu}\frac{\slashed{q}+\slashed{k}_2+m}
{m^2-\left(q+k_2\right)^2}\gamma^{\mu}\right]\right\},
\end{eqnarray}
\begin{eqnarray}\nonumber
\left[{\rm d} u\right]_t\equiv
\frac{{\rm d}^4 u_1}{(2\pi)^4}\frac{{\rm d}^4 u_2}{(2\pi)^4}\Delta_t(u_1)\Delta_t(u_2)\,, \quad
k_1^0=|\bm{k}_1|\,,\quad k_2^0=|\bm{k}_2|\,,\quad k=k_2-k_1.
\end{eqnarray}\noindent It is understood that $q=q(k_1,k_2)$ in these formulas, the arguments of $q$ being suppressed for brevity. To extract the leading large-time contribution, we rewrite Eq.~(\ref{dRo1}) as
\begin{eqnarray}\label{dRo2}
\delta_C\varrho(t,\bm{q})=
\int &&\hspace{-0,3cm}\frac{{\rm d}^3 \bm{k}_1}{(2\pi)^3}\int\frac{{\rm d}^3 \bm{k}_2}{(2\pi)^3}
\int_{-\infty}^{+\infty}{\rm d}\xi_z\int\left[{\rm d} u\right]_{\delta t}\left[{\rm d} v\right]_{\delta t}\delta(k_{1z}-k^{\ast}_{1z})\\&&
\times\frac{(2/\pi)\varepsilon^2_{\bm{q}}w(\bm{q},\bm{k}_1+\bm{\xi},\bm{k}_2)n(\bm{k}_1)[1+n(\bm{k}_2)]
\varrho(t,\bm{q}+\bm{k})}{[m^2-\left(q(k_1+\xi,k_2)+v_1+v_2\right)^2-{\rm i}0][m^2-\left(q(k_1+\xi,k_2)-u_1-u_2\right)^2+{\rm i}0]}\,,\nonumber
\end{eqnarray}
where $\xi = (0,\bm{\xi}),$ $\bm{\xi} = (0,0,\xi_z),$ and $k^{\ast}_{1z}$ is the root of $q^2\left(k_1,k_2\right)=m^2$ with respect to $k_{1z}.$ Indeed, a shift
$\xi_z\to\xi_z-k_{1z}$ followed by integration over $k_{1z}$ removes the $\delta$-function $\delta(k_{1z}-k^{\ast}_{1z}),$ bringing us back to Eq.~(\ref{dRo1}). The leading term comes from integration near $\xi = 0.$ Therefore, when extracting this term, one can set $\xi=0$ in the numerator of the integrand in Eq.~(\ref{dRo2}). The function $w(\bm{q},\bm{k}_1,\bm{k}_2)$ given by Eq.~(\ref{w}) [in which $q(k_1,k_2)$ is now on the mass shell] is then nothing but the probability of the scattering $${\rm electron}(q) + {\rm photon}(k_2) \to {\rm electron}(q+k) + {\rm photon}(k_1),$$ where ``${\rm electron}(q)$'' denotes unpolarized electron with momentum $q,$ and ``${\rm photon}(k)$'' a photon with momentum $k$ in any of the two polarization states over which summation is done for the initial as well as final photons. In the non-relativistic approximation,
\begin{eqnarray}\nonumber
w(\bm{q},\bm{k}_1,\bm{k}_2)=\frac{\pi e^4}{2m^2|\bm{k}_1||\bm{k}_2|}
\left[1+\frac{(\bm{k}_1\bm{k}_2)^2}{\bm{k}_1^2\bm{k}_2^2}\right].
\end{eqnarray}
In the denominator, we expand $q(k_1+\xi,k_2)$ with respect to $\xi$ to the first order
$$m^2-\left(q(k_1+\xi,k_2)+v_1+v_2\right)^2 \to
-2 q^0(k_1,k_2)\left(v^0_1+v^0_2 +\xi_z\frac{\partial q^0(k_1,k_2)}{\partial k_{1z}}\right).$$
Introducing a new integration variable
$\zeta=\xi_z\partial q^0(k_1,k_2)/\partial k_{1z},$ and using
$$\delta(k_{1z}-k^{\ast}_{1z})=
\left|\frac{\partial q^0(k_1,k_2)}{\partial k_{1z}}\right|
\delta\left(q^0(k_1,k_2)-\varepsilon_{\bm{q}}\right)$$
thus gives
\begin{eqnarray}\label{deltac}
\delta_C\varrho(t,\bm{q})=I(\delta t)
\int\frac{{\rm d}^3 \bm{k}_1}{(2\pi)^3}\int\frac{{\rm d}^3 \bm{k}_2}{(2\pi)^3}w(\bm{q},\bm{k}_1,\bm{k}_2)n(\bm{k}_1)[1+n(\bm{k}_2)]
\varrho(t,\bm{q}+\bm{k})
\delta\left(q^0(k_1,k_2)-\varepsilon_{\bm{q}}\right),
\nonumber\\
\end{eqnarray}\noindent
where
$$I(t)=-\frac{1}{2\pi}\int_{-\infty}^{+\infty} {\rm d}\zeta\int
\frac{\left[{\rm d} u\right]_t\left[{\rm d} v\right]_t}{(v^0+\zeta+{\rm i}0)(u^0-\zeta+{\rm i}0)}.$$ Integrations are done with the help of the formulas
$$\int\frac{\left[{\rm d} v\right]_t}{v^0+\zeta+{\rm i}0}
=\int\frac{{\rm d}v^0_1}{2\pi {\rm i}}\int\frac{{\rm d}v^0_2}{2\pi {\rm i}}
\frac{e^{-{\rm i}v^0_1 t}-1}{v^0_1}\frac{e^{-{\rm i}v^0_2 t}-1}{v^0_2}\frac{1}{v^0_1+v^0_2+\zeta+{\rm i}0}
=\frac{1-e^{{\rm i}\zeta t}}{\zeta},$$
$$\int_{-\infty}^{+\infty}{\rm d}\zeta\frac{1-e^{{\rm i}\zeta t}}{\zeta}\frac{1-e^{-{\rm i}\zeta t}}{\zeta} = 2\pi t.$$ The result is $I(t) = t.$ Substituting this into Eq.~(\ref{deltac}), and dividing by $\delta t,$ we find the connected contribution to the derivative $\partial\varrho/\partial t.$ Comparison with Eq.~(\ref{dRo_dt}) now gives (we use $q^0(k_1,k_2)=\varepsilon_{\bm{q}-\bm{k}_1+\bm{k}_2}+|\bm{k}_1|-|\bm{k}_2|,$ and change notation $\bm{k}_1 \to \bm{p},$ so that $\bm{k}_2 = \bm{p}+\bm{k}$)
\begin{eqnarray}
C(\bm{q},\bm{k})=\int\frac{{\rm d}^3 \bm{p}}{(2\pi)^3}w(\bm{q},\bm{p},\bm{k}+\bm{p})
\delta\left(\varepsilon_{\bm{q}+\bm{k}}+|\bm{p}|-|\bm{p}+\bm{k}|-\varepsilon_{\bm{q}}\right)
n(\bm{p})\left[1+n(\bm{p}+\bm{k})\right].\nonumber
\end{eqnarray}\noindent Thus, Eq.~(\ref{dRo_dt}) becomes
\begin{eqnarray}\label{Boltz}
\frac{\partial\varrho(t,\bm{q})}{\partial t}=
&&\hspace{-0,5cm}\int\frac{{\rm d}^3 \bm{k}}{(2\pi)^3}\int\frac{{\rm d}^3 \bm{p}}{(2\pi)^3}
\delta\left(\varepsilon_{\bm{q}+\bm{k}}+|\bm{p}|-|\bm{p}+\bm{k}|-\varepsilon_{\bm{q}}\right)
w(\bm{q},\bm{p},\bm{k}+\bm{p})\nonumber\\&&
\times\left\{n(\bm{p})\left[1+n(\bm{p}+\bm{k})\right]\varrho(t,\bm{q}+\bm{k})
-n(\bm{p}+\bm{k})\left[1+n(\bm{p})\right]\varrho(t,\bm{q})\right\}.
\end{eqnarray}\noindent The right-hand side of this equation is the standard form of the collision integral as representing the difference of ``gains'' and ``losses'' of the electron due to its collisions with photons. Substituting Eq.~(\ref{nk}) for the photon distribution, one finds that the equilibrium electron momentum distribution $\varrho(\bm{q})\sim \exp(-\beta \varepsilon_{\bm{q}}).$ It is to be noted that no quasi-classic condition on the electron state has been imposed in the above derivation. In fact, the existence of irreversible spreading implies that the large-time evolution of the electron cannot be considered quasi-classically.

\section{Discussion and conclusions}\label{conclusions}

As was explained in the introduction, the reason that makes the effective field formalism indispensable in quantum field theory is the principal incompleteness of the S-matrix approach regarding description of the field measurements. A general conclusion of our investigation is that the presence of infrared divergences in the field expectation values is not a sign of restricted validity of the effective field formalism, but rather an indication on inadequacy of the standard approach based on the assumption that the problem admits an infinite temporal extension. Specifically, we have shown that restricting consideration to a finite time interval makes the effective field infrared-finite, and that the presence of infrared divergences in the standard approach means that the radiative corrections to the classical field grow unboundedly with time at every order of perturbation theory. We proved also that the sum of these contributions is bounded, and is such that the effective electromagnetic field of an electron freely moving in a photon bath vanishes in the large-time limit. As the simple example given in Sec.~\ref{gausswave} demonstrates, this effect is irreversible -- the electron coordinate variance cannot be made less than $\sim \alpha T t/m^2,$ if the electron travels for a time $t$ in a photon bath at temperature $T.$ This conclusion holds true also in vacuum, though the effect is much weaker in that case. Thus, the physical meaning of the infrared singularity in the effective electromagnetic field is the existence of irreversible spreading of electric charges. This conclusion was formulated in Ref.~\cite{jpa} as a natural interpretation of the results obtained using the momentum-cutoff regularization in the standard approach. Now that essentially the same results have been obtained directly from the fundamental principles of quantum field theory, without having to introduce an auxiliary infrared regularization and to use an {\it ad hoc} definition of the effective density matrix through the effective field, the existence of irreversible spreading  is proved to be an unequivocal consequence of quantum electrodynamics.

It should be stressed that the irreversible spreading does not affect the scattering cross-sections themselves. As long as the single electron evolution in a photon bath is considered, this follows directly from Eq.~(\ref{diag}) and the infrared finiteness of  diagonal elements of the electron density matrix, proved in Sec.~\ref{factorization}. But this result equally applies to any scattering process. The point is that the scattering amplitudes can be constructed entirely in terms of momenta (and polarizations) of the free particles present in the initial and final states; the standard procedure is to formally replace particle wave-packets by infinitely wide homogeneous beams of identical particles, erasing thereby any information about spatial profiles of actual particle states. In other words, it is diagonal elements of the momentum density matrices of particles that only matter when computing the cross-sections, and these are unaffected by the infrared singularity.

Next, some technical remarks are in order. First of all, regarding the strength of the irreversible spreading, the role of four-dimensionality of spacetime must be emphasized. That this factor is crucial is evident already from the relation of this spreading to the infrared singularities of radiative corrections, but is particularly clear from comparison with the usual electron-photon scattering responsible for the electron momentum relaxation considered in Sec.~\ref{relaxation}. The latter is also described by the loop contributions which diverge for $t\to \infty$ (see Fig.~\ref{heat}), but by those only which are due to integration over finite photon momenta, and which are therefore insensitive to the spacetime dimensionality. Furthermore, it is the four-dimensionality of spacetime that eventually allows an infinitesimal treatment of the problem, given in Sec.~\ref{effmat_calc}. We note in this connection that the results of Sec.~\ref{effmat_calc} provide an effective tool for investigating the infrared problem in non-Abelian gauge theories.

Comparison of the two stages of the electron evolution is useful also when interpreting the infrared singularity as a thermalization of the electron state. The infrared thermalization turns out to be stronger than that effected by the electron-photon scattering, in two respects. The infrared damping of the off-diagonal elements of the matrix $\varrho(t;\bm{q},\bm{q}')$ is more complete, as it takes place for all $\bm{q}' \ne \bm{q},$ whereas the usual thermalization implies only vanishing of the matrix elements between states with different energy, that is, $\bm{q}'^2 \ne \bm{q}^2.$ The other essential difference is that the infrared thermalization is an $O(\alpha)$-effect, whereas the electron-photon collision effects are $O(\alpha^2),$ which is reflected in the ratio of characteristic times of the two stages (see Eq.~(\ref{tauratio})). The role of temperature in these processes is also quite revealing: in contrast to the usual relaxation, the infrared thermalization takes place in vacuum as well as at $T\ne 0.$ All these distinctions accentuate the peculiar nature of this phenomenon which has no proper analogy in nonrelativistic physics.

\acknowledgments
We thank F.~A.~Egorov (Russian Academy of Science), P.~Pronin, A.~Baurov and all participants of the theoretical physics seminar at the Physics Faculty of Moscow State University for their interest to our work and useful discussions.

\begin{appendix}
\section{Gauss law in Lorentz gauge}
\begin{figure}[b]
\begin{center}
     \includegraphics{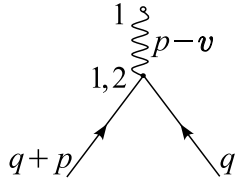}
\caption{The tree contribution to the effective electromagnetic field of electron.}
\label{drev}
\end{center}
\end{figure}
We saw in Sec.~\ref{form} that the use of covariant techniques to compute the mean field is hindered by the fact that on a finite time interval, the usual procedure of adiabatic switching of the interaction is not applicable, which leads to appearance of a non-covariant term in the interaction Lagrangian (the last term in Eq.~(\ref{L_l})). In this appendix, we demonstrate on a simple example that this complication is not a mere formality: omission of this term leads to violation of the Gauss law already in the tree approximation. In this approximation, the mean electromagnetic field is represented by the diagrams in Fig.~\ref{drev}, whose analytic expression is
\begin{eqnarray}\label{amu_g}
A_\mu^{\rm eff}(t,\bm{x})= - e\int\frac{{\rm d}^3 \bm{p}}{(2\pi)^3}\int\frac{{\rm d}^4 v}{(2\pi)^4}&&\hspace{-0,3cm}
J^\alpha(\bm{p}){\rm e}^{{\rm i}(p_0-v_0)t-{\rm i}\bm{p}\bm{x}}
\overline{\Delta}_t(v)g^\nu_\alpha(p-v)\nonumber\\&&
\times\left[D^{(11)}_{\mu\nu}(p-v)-D^{(21)}_{\mu\nu}(p-v)\right],
\end{eqnarray}\noindent where
\begin{eqnarray}\nonumber
J^\alpha(\bm{p})=\int\frac{{\rm d}^3 \bm{q}}{(2\pi)^3}\sum_{\sigma\sigma'}
\varrho_{\sigma\sigma'}(\bm{q},\bm{q}+\bm{p})
\frac{\bar{u}_{\sigma'}(\bm{q}+\bm{p})\gamma^\alpha u_{\sigma}(\bm{q})}
{\sqrt{2\varepsilon_{\bm{q}+\bm{p}}}\sqrt{2\varepsilon_{\bm{q}}}}\,.
\end{eqnarray}\noindent
As was explained in Sec.~\ref{rules}, we may replace here $d_{\mu\nu}$ by $\eta_{\mu\nu},$ so that
\begin{eqnarray}\nonumber
D^{(11)}_{\mu\nu}(k)-D^{(21)}_{\mu\nu}(k)=
\left[\frac{1}{k^2+{\rm i}0}+2\pi{\rm i}\theta(k_0)\delta(k^2)\right]\eta_{\mu\nu}.
\end{eqnarray}\noindent
Substituting this into Eq.~(\ref{amu_g}), performing the integration over $v$ by closing the contour of $v^0$-integration in the lower half-plane ($t>0$), and using the identity $p_\mu J^\mu(\bm{p})=0$ we obtain
\begin{eqnarray}\label{a_mu_fin}
A_\mu^{\rm eff}(t,\bm{x})=e\int\frac{{\rm d}^3 \bm{p}}{(2\pi)^3}
{\rm e}^{-{\rm i}\bm{p}\bm{x}}
\left[-J_\mu(\bm{p})\left\{\frac{{\rm e}^{{\rm i}p_0 t}}{p^2}
-\frac{1}{2|\bm{p}|}\left(\frac{{\rm e}^{{\rm i}|\bm{p}|t}}{p_0-|\bm{p}|}-
\frac{{\rm e}^{-{\rm i}|\bm{p}|t}}{p_0+|\bm{p}|}\right)\right\}\right.\nonumber\\
\left.
+J^0(\bm{p})\frac{p_0\eta_\mu-p_{\mu}}{\bm{p}^2}
\frac{{\rm e}^{{\rm i}|\bm{p}|t}-{\rm e}^{-{\rm i}|\bm{p}|t}}{2|\bm{p}|}
\right].
\end{eqnarray}\noindent The term proportional to $J^0(\bm{p})$ is the contribution of the non-covariant term in $L_l.$ Differentiation of Eq.~(\ref{a_mu_fin}) gives
\begin{eqnarray}\nonumber
\partial^\mu A_\mu^{\rm eff}(t,\bm{x})=0\,, \quad
\Box A_\mu^{\rm eff}(t,\bm{x})=eJ_\mu(t,\bm{x})\,,\quad {\rm where}\quad
J_\mu(t,\bm{x})=\int\frac{{\rm d}^3 \bm{p}}{(2\pi)^3}
{\rm e}^{{\rm i}p x}J_\mu(\bm{p}).
\end{eqnarray}\noindent
Suppose now that the non-covariant term is omitted (which is equivalent to replacing $g^\nu_\alpha\to\delta^\nu_\alpha$ in Eq.~(\ref{amu_g})). Then the equation $\Box A_\mu^{\rm eff}=eJ_\mu$ still holds, but
\begin{eqnarray}\label{div_a}
\partial^\mu A_\mu^{\rm eff}(t,\bm{x}) =
e\int\frac{{\rm d}^3 \bm{p}}{(2\pi)^3}
J_0(\bm{p})\frac{{\rm e}^{-{\rm i}\bm{p}\bm{x}}\sin|\bm{p}|t}{|\bm{p}|}\,.
\end{eqnarray}\noindent
It is not difficult to see that the latter equation is inconsistent with the Gauss law. For instance, consider a heavy particle localized in a small vicinity of the origin, so that  $\bm{J}\approx 0,$ $J_0(t,\bm{x})\approx \delta^{(3)}(\bm{x}).$ Then integration of Eq.~(\ref{div_a}) and $\Box A_\mu^{\rm eff}=eJ_\mu$ gives
\begin{eqnarray}
\bm{A}^{\rm eff}(t,\bm{x})=0\,,\quad
A_0^{\rm eff}(t,\bm{x})=\frac{e}{4\pi r}\theta(t-r)\,,
\end{eqnarray}\noindent
where $r$ is the distance between the charge and the observation point. Thus, at any given distance $r,$ the Gauss law holds only at times $t>r.$ In general, this law is restored only asymptotically: it is seen from Eq.~(\ref{div_a}) that because of the factor $\sin|\bm{p}|t,$ $\partial^{\mu}A_\mu^{\rm eff}$ exponentially tends to zero as $t\to \infty.$

In the tree approximation, this difficulty with the charge conservation can be overcome by imposing some special conditions on the photon state. However, omission of the non-invariant term in Eq.~(\ref{L_l}) turns out to be much more harmful in higher orders of perturbation theory, namely, it leads to gauge-dependence of the effective current, which cannot be cured by modifying the photon state vector.

\end{appendix}

\end{document}